# First M87 Event Horizon Telescope Results. II.
# Array and Instrumentation

The Event Horizon Telescope Collaboration
(See the end matter for the full list of authors.)

## Abstract

The Event Horizon Telescope (EHT) is a very long baseline interferometry (VLBI) array that comprises millimeter- and submillimeter-wavelength telescopes separated by distances comparable to the diameter of the Earth. At a nominal operating wavelength of ~1.3 mm, EHT angular resolution ($\lambda/D$) is ~25 $\mu$as, which is sufficient to resolve nearby supermassive black hole candidates on spatial and temporal scales that correspond to their event horizons. With this capability, the EHT scientific goals are to probe general relativistic effects in the strong-field regime and to study accretion and relativistic jet formation near the black hole boundary. In this Letter we describe the system design of the EHT, detail the technology and instrumentation that enable observations, and provide measures of its performance. Meeting the EHT science objectives has required several key developments that have facilitated the robust extension of the VLBI technique to EHT observing wavelengths and the production of instrumentation that can be deployed on a heterogeneous array of existing telescopes and facilities. To meet sensitivity requirements, high-bandwidth digital systems were developed that process data at rates of 64 gigabit s$^{-1}$, exceeding those of currently operating cm-wavelength VLBI arrays by more than an order of magnitude. Associated improvements include the development of phasing systems at array facilities, new receiver installation at several sites, and the deployment of hydrogen maser frequency standards to ensure coherent data capture across the array. These efforts led to the coordination and execution of the first Global EHT observations in 2017 April, and to event-horizon-scale imaging of the supermassive black hole candidate in M87.

## 1. Introduction

It is generally accepted that active galactic nuclei (AGNs) are powered by accretion onto supermassive black holes (SMBHs; Heckman & Best 2014). These central engines are powerful actors on the cosmic stage, with roles in galactic evolution, star formation, mergers, and particle acceleration as evidenced by relativistic jets that both dynamically influence and redistribute matter on galactic scales (Blandford et al. 2018). Inflowing material typically obscures the event horizons of these black hole candidates, but it is in this extreme environment of the black hole boundary that strong-field effects of general relativity become evident and the accretion and outflow processes that govern black hole feedback on galactic scales originate (Ho 2008). Imaging black holes on scales that resolve these effects and processes would enable new tests of general relativity and the extraordinarily detailed study of core AGN physics. Realization of this goal requires a specialized instrument that does two things. It must have the ultra-high angular resolution required to resolve the nearest SMBH candidates, and it must operate in a range of the electromagnetic spectrum where light streams unimpeded from the innermost accretion region to telescopes on Earth. Achieving these specifications is the primary objective of the Event Horizon Telescope (EHT): a very long baseline interferometry (VLBI) array of millimeter (mm) and submillimeter (submm) wavelength facilities that span the globe, creating a telescope with an effective Earth-sized aperture (Doeleman et al. 2009a).

While the EHT is uniquely designed for the imaging of SMBHs, other pioneering instruments are capable of probing similar angular scales for other purposes. Operating in the infrared, the GRAVITY interferometer delivers relative astrometry at the ~10 micro-arcsecond ($\mu$as) level and has provided evidence for relativistic motion of material in close proximity to Sgr A* (GRAVITY Collaboration et al. 2018a). These infrared observations are an important and parallel probe of the spacetime surrounding Sgr A*, but cannot be used to make spatially resolved images of the black hole candidate because the interferometer intrinsic resolution is only 3 mas. The RadioAstron satellite, used as an orbiting element of combined Earth-Space VLBI arrays, is also capable of ~10 $\mu$as angular resolution (Kardashev et al. 2013), but it operates at longer radio wavelengths that cannot penetrate the self-absorbed synchrotron plasma that surrounds the event horizon. In many ways the EHT is complementary to the Laser Interferometer Gravitational-Wave Observatory (LIGO) facility, which has detected the gravitational wave signatures from merging stellar-mass black holes (Abbott et al. 2016). LIGO and the EHT observe black holes that differ in mass by factors of $10^4$–$10^7$; LIGO events are transient, while the EHT carries out long-term studies of its main targets.

This Letter is one in a sequence of manuscripts that describes the first EHT results. The full sequence includes an abstract Letter with a summary of results (EHT Collaboration et al. 2019a, hereafter Paper I), this Letter with a description of the array and instrumentation (Paper II), a description of the data pipeline and processing (EHT Collaboration et al. 2019b, hereafter Paper III), a description of imaging techniques (EHT Collaboration et al. 2019c, hereafter Paper IV), and theoretical analyses of astrophysical and physics results (EHT

Collaboration et al. 2019d, hereafter Paper V; EHT Collaboration et al. 2019e, hereafter Paper VI, respectively).

### 1.1. EHT Science Goals

The scientific goals of the EHT array are to resolve and detect strong general relativistic signatures expected to arise on event-horizon scales. The best-known effect is that a black hole, surrounded by an optically thin luminous plasma, should exhibit a "silhouette" or "shadow" morphology: a dim central region delineated by the lensed photon orbit (Falcke et al. 2000). The apparent size of the photon orbit, described not long after Schwarzschild's initial solution was published (Hilbert 1917; von Laue 1921), defines a bright ring or crescent shape that was calculated for arbitrary spin by Bardeen (1973), first imaged through simulations by Luminet 1979, and subsequently studied extensively (Chandrasekhar 1983; Takahashi 2004; Broderick & Loeb 2006). The size and shape of the resulting shadow depends primarily on the mass of the black hole, and only very weakly on its spin and the observing orientation. For a non-spinning black hole, the shadow diameter is equal to $\sqrt{27}$ Schwarzschild radii ($R_s = 2GM/c^2$). Over all black hole spins and orientations, the shadow diameter ranges from 4.8 to $5.2\,R_s$ (Bardeen 1973; Johannsen & Psaltis 2010). Because any light that crosses the photon orbit from outside will eventually reach the event horizon, use of the term "horizon scale" will hereafter be understood to mean the size of the shadow and the lensed photon orbit. Detection of the shadow via lensed electromagnetic radiation would provide new evidence for the existence of SMBHs by confining the masses of EHT targets to within their expected photon orbits. A more detailed study of the precise shape of the photon orbit can be used to test the validity of general relativity on horizon scales (Johannsen & Psaltis 2010). Full polarimetric imaging can similarly be used to map magnetic field structure near the event horizon, placing important constraints on modes of accretion and the launching of relativistic jets (Broderick & Loeb 2009; Johnson et al. 2015; Chael et al. 2016; Akiyama et al. 2017; Gold et al. 2017).

Separate signatures, potentially offering a more sensitive probe of black hole spin, are the timescales of dynamical processes at horizon scales. A characteristic timescale for such processes is given by the orbital period of test particles at the innermost stable circular orbit (ISCO), which depends sensitively on the black hole spin (Bardeen et al. 1972). Monitoring VLBI observables to track the orbital dynamics of inhomogeneities in the accretion flow can thus be used to probe the spacetime, and potentially spin, of the black hole (Broderick & Loeb 2006; Doeleman et al. 2009b; Fish et al. 2009; Fraga-Encinas et al. 2016; Medeiros et al. 2017; Roelofs et al. 2017).

### 1.2. Target Sources and Confirmation of Horizon-scale Structure

The bright radio core of M87 and Sgr A* (Table 1) are the primary EHT targets, as the combination of their estimated mass and proximity make them the two most suitable sources for studying SMBH candidates at horizon-scale resolution (Narayan & McClintock 2008; Johannsen et al. 2012). With bolometric luminosities well below the Eddington limit (Di Matteo et al. 2000; Baganoff et al. 2003), both the nucleus of M87 and Sgr A* are representative of the broad and populous class of low-luminosity AGN (LLAGN). AGN spend most of their time in this low state after prior periods of high accretion (Ho 2008), and LLAGN share many characteristics of stellar black hole emission in X-ray binaries, which exhibit episodic emission and jet production (Narayan & McClintock 2008). Thus, the prospects for applying results from horizon-scale observations of the central region of M87 and Sgr A* across a broad range of astrophysical contexts are excellent.

Within the past decade, VLBI observations at a wavelength of 1.3 mm have confirmed the existence of structure on the scale of the shadow in both the nucleus of M87 and Sgr A*. M87 (Virgo A) is thought to harbor a black hole in the range of $3.3–6.2 \times 10^9$ solar mass ($M_\odot$) (Gebhardt et al. 2011; Walsh et al. 2013), when scaled to a distance of 16.8 Mpc (Paper VI). Structure in M87 measuring 40 $\mu$as in extent has been resolved with 1.3 mm VLBI, corresponding to a total extent of ~5.5 Schwarzschild radii at the upper end of the mass range (Doeleman et al. 2012; Akiyama et al. 2015). For Sgr A*, the black hole candidate at the Galactic center with a presumed mass of $4 \times 10^6\,M_\odot$ (Balick & Brown 1974; Ghez et al. 2008; Genzel et al. 2010; GRAVITY Collaboration et al. 2018a), the 1.3 mm emission has been measured to have a scale of $3.7\,R_s$ (Doeleman et al. 2008; Fish et al. 2011). More recently, full polarimetric VLBI observations at 1.3 mm wavelength have revealed ordered and time-variable magnetic fields within Sgr A* on horizon scales (Johnson et al. 2015), and extension to longer baselines has confirmed compact structure on ~3 $R_s$ scales (Lu et al. 2018). These results, obtained with three- and four-site VLBI arrays consisting of the former Combined Array for Research in Millimeter-wave Astronomy (CARMA) in California, the Submillimeter Telescope (SMT) in Arizona, the James Clerk Maxwell Telescope (JCMT) and Submillimeter Array (SMA) facilities on Maunakea in Hawaii, and the Atacama Pathfinder Experiment (APEX) telescope in Chile, demonstrated that direct imaging of emission structures near the event horizon of SMBH candidates is possible in principle. For comparison and perspective, the closest approach of the orbiting stars used to determine the mass of Sgr A* is ~1400 $R_s$ (Gravity Collaboration et al. 2018b).

### 1.3. Array Architecture and Context

To realize these fundamental science goals, our international collaboration has engineered the EHT to move beyond the detection of horizon-scale structure and achieve the required imaging and time-domain sampling capability.

One of the key enabling technologies behind the EHT observations has been the development of high-bandwidth (wideband) VLBI systems that compensate to some degree for the generally smaller telescope apertures at millimeter and submillimeter wavelengths. The first detections of horizon-scale structure followed directly from deployment of new digital VLBI backend and recording instrumentation, custom-built for mm-wavelength observations that achieved a recording rate of 4 gigabit s$^{-1}$ (Gbps)[136] (Doeleman et al. 2008). Continued development led to an increased recording rate of 16 Gbps (Whitney et al. 2013). Adoption of industry-standard high-speed data protocols, increased hard disk storage capacity,

---

[136] The bandwidth of an EHT observation is typically expressed as a recording rate in gigabit-per-second, or Gbps. The recording rate in Gbps is four times the recorded bandwidth in GHz, from a factor of two coming from the need to take samples at a rate of twice the bandwidth (Nyquist rate), and another factor of two because each sample is 2 bits.

**Table 1**
Assumed Physical Properties of Sgr A* and M87 Used to Establish Technical Goals[a]

| | | Sgr A* | M87 |
|---|---|---|---|
| Black Hole Mass | $M$ ($M_\odot$) | $4.1 \times 10^6$ (1) | $(3.3-6.2) \times 10^9$ (5), (6) |
| Distance | $D$ (pc) | $8.34 \times 10^3$ (2) | $16.8 \times 10^6$ (7) |
| Schwarzschild Radius | $R_s$ ($\mu$as) | 9.7 | 3.9–7.3 |
| Shadow Diameter[b] | $D_{sh}$ ($\mu$as) | 47–50 | 19–38 |
| Brightness Temperature[c] | $T_B$ (K) | $3 \times 10^9$ (3) | $10^{10}$ (8) |
| Period ISCO[d] | $P_{ISCO}$ | 4–54 minutes | 2.4–57.7 days |
| Mass Accretion Rate[e] | $\dot{M}$ ($M_\odot$ yr$^{-1}$) | $10^{-9}-10^{-7}$ (4) | $< 10^{-3}$ (9) |

**Notes.**
[a] Sgr A*: $\alpha_{J2000.0} = 17^h45^m40^s.0409$, $\delta_{J2000.0} = -29°00'28''.118$ (10); M87: $\alpha_{J2000.0} = 12^h30^m49^s.4234$, $\delta_{J2000.0} = 12°23'28''.044$ (11).
[b] The shadow diameter is within the range $4.8-5.2 R_s$ depending on black hole spin and orientation to the observer's line of sight (Johannsen & Psaltis 2010).
[c] Brightness temperatures are reported for an observing frequency of 230 GHz.
[d] $P_{ISCO}$ range is given in the case of maximum spin for both prograde (shortest) and retrograde (longest) orbits (Bardeen et al. 1972).
[e] Mass accretion rates $\dot{M}$ are estimated from measurements of Faraday rotation imparted by material in the accretion flow around the black hole.

**References.** (1) GRAVITY Collaboration et al. (2018a), (2) Reid et al. (2014), (3) Lu et al. (2018), (4) Marrone et al. (2007), (5) Walsh et al. (2013), (6) Gebhardt et al. (2011), (7) Blakeslee et al. (2009), EHT Collaboration et al. (2019e), (8) Akiyama et al. (2015), (9) Kuo et al. (2014), (10) Reid & Brunthaler (2004), (11) Lambert & Gontier (2009).

and flexible field programmable gate array (FPGA) computational fabric have enabled the EHT to reach data throughputs of 64 Gbps (Vertatschitsch et al. 2015), or 32 times the maximum recording rate and corresponding bandwidth, offered by open access VLBI facilities at longer wavelengths (e.g., the NRAO Very Long Baseline Array; Napier et al. 1994).

In addition to the increased sensitivity provided by such large data recording rates, several factors, some engineered and some serendipitous, have converged to enable spatially and temporally resolved observations of black hole candidates by the EHT. By temporal resolution we mean that increased resolution, sensitivity, and baseline coverage allows the EHT to detect and spatially resolve horizon-scale time-variable structures, which would otherwise only be studied through light-curve analysis and light-crossing time assumptions. A list of the key enabling factors is as follows.

1. *Angular resolution*: the angular resolution of Earth-diameter VLBI baselines at wavelengths of ~1.3 mm can resolve the lensed photon orbits of Sgr A* and M87 (Table 1: about 50 $\mu$as and 38 $\mu$as, respectively).
2. *Fourier coverage*: using Earth-rotation aperture synthesis, the number of existing and planned mm/submm wavelength telescopes provides a sufficient sampling of VLBI baseline lengths and orientations to produce images with horizon-scale resolution (Fish et al. 2014; Honma et al. 2014; Lu et al. 2014; Ricarte & Dexter 2015; Bouman et al. 2016; Chael et al. 2016).
3. *Atmospheric transparency*: at the required mm/submm observing wavelengths, the Earth's atmosphere at high-altitude sites is reliably transparent enough that global VLBI arrays can be formed for long-duration observations (Thompson et al. 2017).
4. *Optically thin accretion*: for both Sgr A* and M87 the spectral energy density of the accretion flow begins to turn over at mm wavelengths, allowing photons to escape, presuming a synchrotron emission mechanism (see Broderick & Loeb 2009; Genzel et al. 2010, for M87 and Sgr A*, respectively).

5. *Interstellar scattering*: radio images of Sgr A* are blurred due to interstellar scattering by free electrons (Lo et al. 1998; Shen et al. 2005; Bower et al. 2006; Lu et al. 2011; Bower et al. 2015; Johnson et al. 2018; Psaltis et al. 2018). This blurring decreases with wavelength as $\lambda^2$ and becomes sub-dominant for wavelengths of ~1.3 mm and shorter, where observations enable direct access to intrinsic structures in close proximity to the event horizon (Doeleman et al. 2008; Issaoun et al. 2019).

### 1.4. Current EHT Array

Figure 1 shows a map of the EHT array. In 2017 April, the EHT carried out global observations with an array of eight telescopes (see Table 2) that included the Atacama Large Millimeter/submillimeter Array (ALMA) for the first time. A purpose-built system electronically combined the collecting area of ~37 × 12 m diameter ALMA dishes (see Appendix A.1): the equivalent of adding a ~70 m dish to the EHT array. Other participating telescopes were APEX, JCMT, SMA, SMT, the Large Millimeter Telescope Alfonso Serrano (LMT), the Pico Veleta 30 m telescope (PV), and the South Pole Telescope (SPT). Operating in the 1.3 mm window in full polarimetric mode and with an aggregate bandwidth of 8 GHz, the resulting increase in sensitivity above the first horizon-scale detections was nearly an order of magnitude (e.g., Section 3.8). For observations with phased-ALMA in 2018, the EHT added an additional facility (the 12 m diameter Greenland Telescope (GLT)) and doubled the aggregate bandwidth to its nominal target of 16 GHz.

The sections that follow describe the specifications and characteristics of the array (Section 2), the EHT instrumentation deployed (Section 3), the observing strategy (Section 4), correlation, calibration, and detection (Section 5), and future enhancements (Section 6).

## 2. EHT Specifications and Characteristics

Extending the VLBI technique to wavelengths of ~1 mm presents technical challenges. Heterodyne receivers exhibit

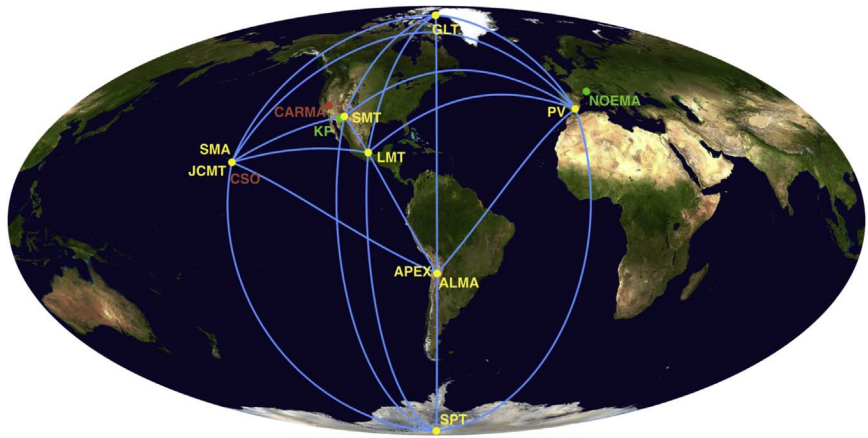

**Figure 1.** Map of the EHT. Stations active in 2017 and 2018 are shown with connecting lines and labeled in yellow, sites in commission are labeled in green, and legacy sites are labeled in red. Nearly redundant baselines are overlaying each other, i.e., to ALMA/APEX and SMA/JCMT. Such redundancy allows improvement in determining the amplitude calibration of the array (Paper III).

**Table 2**
EHT Station Information

| Facility | Diameter (m) | Location | $X$[a] (m) | $Y$[a] (m) | $Z$[a] (m) | Latitude | Longitude | Elevation[a] (m) |
|---|---|---|---|---|---|---|---|---|
| | | | Facilities that Participated in the 2017 Observations | | | | | |
| ALMA[b] | 12 (×54) and 7 (×12) | Chile | 2225061.3 | −5440061.7 | −2481681.2 | −23°01′45″1 | −67°45′17″1 | 5074.1 |
| APEX | 12 | Chile | 2225039.5 | −5441197.6 | −2479303.4 | −23°00′20″8 | −67°45′32″9 | 5104.5 |
| JCMT | 15 | Hawaii, USA | −5464584.7 | −2493001.2 | 2150654.0 | +19°49′22″2 | −155°28′37″3 | 4120.1 |
| LMT | 50 | Mexico | −768715.6 | −5988507.1 | 2063354.9 | +18°59′08″8 | −97°18′53″2 | 4593.3 |
| PV 30 m | 30 | Spain | 5088967.8 | −301681.2 | 3825012.2 | +37°03′58″1 | −3°23′33″4 | 2919.5 |
| SMA[b] | 6 (×8) | Hawaii, USA | −5464555.5 | −2492928.0 | 2150797.2 | +19°49′27″2 | −155°28′39″1 | 4115.1 |
| SMT | 10 | Arizona, USA | −1828796.2 | −5054406.8 | 3427865.2 | +32°42′05″8 | −109°53′28″5 | 3158.7 |
| SPT[c] | 10 | Antarctica | 809.8 | −816.9 | −6359568.7 | −89°59′22″9 | −45°15′00″3 | 2816.5 |
| | | | Facilities Joining EHT Observations in 2018 and Later | | | | | |
| GLT | 12 | Greenland | 541547.0 | −1387978.6 | 6180982.0 | +76°32′06″6 | −68°41′08″8 | 89.4 |
| NOEMA[d] | 15 (×12) | France | 4524000.4 | 468042.1 | 4460309.8 | +44°38′01″2 | +5°54′24″0 | 2617.6 |
| KP 12 m[d] | 12 | Arizona, USA | −1995954.4 | −5037389.4 | 3357044.3 | +31°57′12″0 | −111°36′53″5 | 1894.5 |
| | | | Facilities Formerly Participating in EHT Observations | | | | | |
| CARMA | 10.4, 6.1 (×8) | California, USA | −2397378.6 | −4482048.7 | 3843513.2 | +37°16′49″4 | −118°08′29″9 | 2168.9 |
| CSO | 10 | Hawaii, USA | −5464520.9 | −2493145.6 | 2150610.6 | +19°49′20″9 | −155°28′31″9 | 4107.2 |

**Notes.**
[a] Geocentric coordinates with $X$ pointing to the Greenwich meridian, $Y$ pointing 90° away in the equatorial plane (eastern longitudes have positive $Y$), and positive Z pointing in the direction of the North Pole. This is a left-handed coordinate system. Elevations are relative to the GRS 80 ellipsoid (Moritz 2000).
[b] Array coordinates indicate the phasing center used in 2017.
[c] 2017 April position: the ice sheet at the South Pole moves at a rate of about 10 m yr$^{-1}$. Effects of this slow drift are removed during VLBI correlation.
[d] NOEMA: Northern Extended Millimeter Array; KP 12 m: Kitt Peak 12 m. These stations have not participated in global EHT observations and their coordinates are approximate.

greater noise, and the required stability of atomic frequency standards is higher than that typically specified for VLBI at longer wavelengths. Early ~1.3 mm wavelength VLBI experiments and observations in the 1990s succeeded in making first detections of AGNs and Sgr A* on modest-length baselines (≲1100 km; Padin et al. 1990; Greve et al. 1995; Krichbaum et al. 1998). Following this pioneering work (see Doeleman & Krichbaum 1999; Boccardi et al. 2017, for summaries), ~1.3 mm VLBI efforts over the next decade focused on order-of-magnitude bandwidth expansion as a means to boost sensitivity. This development led to observations in 2007 with enough resolution to resolve emission on the scale of the event horizon of Sgr A* by using a three-station array with telescopes in Hawaii, California, and Arizona. Motivated by this detection, an EHT array capable of imaging strong general relativistic features was planned based on organizing coordinated observations on a network of mm-wavelength observatories (Doeleman et al. 2009a). These sites and their general characteristics are given in Table 2, which lists current and planned EHT sites as well as now-decommissioned observatories used in prior experiments. For the most part, the EHT consists of pre-existing telescopes that perform single-dish or connected-element astronomy observations during most of the year, but which required modifications or upgrades (in some cases significant ones) to carry out VLBI. The GLT, for example, was commissioned primarily to image M87 (Inoue et al. 2014), and a heterodyne receiver for the SPT was built specifically for EHT observations (Kim et al. 2018a). Details of how each site was modified for EHT work are given in the Appendix.

Technical specifications for the EHT were adopted to unify VLBI recordings across the heterogeneous array by establishing a common frequency configuration, polarization configuration, and sampling rate (Tilanus 2013; Marrone et al. 2014). When ALMA participates in EHT observations, it is by far the most sensitive site, so the overall sensitivity of the EHT array is optimized on a per-bandwidth basis when all other sites can match the recorded frequency bands at ALMA. The EHT converged on a scheme that matches ALMA specifications: two 4 GHz sidebands in each of two polarizations for the 1.3 and 0.87 mm receiver bands, which could be realized through feasible modifications and enhancements at most sites. The resulting global array geometry and sensitivity is well matched to the science goals.

### 2.1. Angular Resolution

Imaging a black hole shadow requires several resolution elements in each direction across the lensed innermost photon orbit, in addition to field-of-view coverage that extends beyond the feature. The longest baselines of the EHT (e.g., South Pole to Arizona, Hawaii, or Spain) provide nominal angular resolutions of $\lambda/D \simeq 25$ $\mu$as in the 1.3 mm wavelength band. Regularized maximum likelihood (RML) imaging methods (see Paper IV) typically achieve angular resolutions that are better than the nominal figure by factors of 2–3 (Narayan & Nityananda 1986). For the EHT case in particular, RML methods have been extensively tested using realistic and synthetic interferometric data to set the optimal resolution of the array (Honma et al. 2014; Bouman et al. 2016; Chael et al. 2016; Akiyama et al. 2017; Chael et al. 2018; Kuramochi et al. 2018). This results in an anticipated effective EHT angular resolution at 1.3 mm of 20 $\mu$as, yielding between about 36 resolution elements across a $120 \times 120$ $\mu$as field of view. For Sgr A* and M87 this field of view is expected to encompass the dim interior, bright annulus of the photon orbit, and sufficient spatial extent to extract the shadow feature (e.g., Mościbrodzka et al. 2009; Dexter et al. 2010). These considerations, and the expectation based on simulations that the EHT instrument and array could achieve such resolution, were important factors in specifying the EHT architecture. Further tests using EHT observations of quasar calibrator sources (e.g., 3C 279) obtained in the 2017 April observations demonstrate robust structural agreement between RML methods and more traditional CLEAN-based radio imaging techniques. Similar comparisons and results on M87, one of the EHT primary targets, can be found in Paper IV.

### 2.2. Sensitivity

Maximizing detections across sparse Fourier coverage is essential for high-fidelity image reconstruction. This is especially true for Sgr A* because its structure is scatter broadened by interstellar effects resulting in reduced VLBI visibility amplitudes on the longest baselines.

To maximize the number of interferometric detections across the array, the EHT uses a two-stage approach to fringe detection. First, detections are found on baselines from all stations to ALMA within the atmospheric coherence time. Next, these ALMA detections are used to remove the effects of phase fluctuations (due to atmospheric turbulence) above the non-ALMA stations, allowing coherent integration on non-ALMA baselines for intervals up to many minutes, thereby boosting the signal-to-noise ratio (S/N) to recover the full baseline coverage of the array (Paper III). Thus, the EHT sensitivity specification corresponds to the requirement that for baselines connecting each EHT site to ALMA, Sgr A* and M87 are detected with a typical signal-to-noise that enables this atmospheric phase correction with acceptable loss. In the usual case where the VLBI observing scan duration greatly exceeds the atmospheric coherence time, the loss due to noise in the phase-correction algorithm is $\sim e^{-(S/N)^2/2}$, where the S/N is the signal-to-noise on the ALMA baselines. To ensure negligible loss, we specify S/N > 3 for EHT detections to ALMA within an integration time, $T_{int}$, where $T_{int}$ is less than the atmospheric coherence time. We note that the expected change in interferometric phase due to source structure over the imaged field of view (Section 2.1) would be less than a few degrees over typical VLBI scan lengths of a few minutes.

The S/N of a VLBI signal on a single baseline between stations is

$$S/N_{(1,2)} = \frac{\eta_Q \sqrt{2\Delta\nu T_{int}} \ S_{cor}}{\sqrt{SEFD_1 \ SEFD_2}} \qquad (1)$$

where $\eta_Q$ is the digital loss due to sampling the received signal at each antenna with finite precision ($\eta_Q = 0.88$ for 2-bit samples), $\Delta\nu$ is the bandwidth of the recording, $SEFD_i$ is the system equivalent flux density (SEFD)[137] for station $i$, and $S_{cor}$ is the expected correlated flux on the baseline between stations 1 and 2. $T_{int}$ is the integration interval of the VLBI signal. It is typically much less than the atmospheric coherence time ($T_{coh}$), or the coherent integration time beyond which the VLBI visibility[138] signal decreases by 10 % due to phase fluctuations imposed by turbulence in the troposphere (typically from a few to $\sim$20 s). On the weakest baseline involving ALMA, ALMA–SPT, with an estimated flux for Sgr A* of 0.1 Jy (see APEX–CARMA baseline in Lu et al. 2018) and with $\Delta\nu = 4$ GHz, an integration time of 3 s yields an S/N of about 12 (see Table 3 for parameters), which far exceeds the EHT S/N specification for detections on ALMA baselines. Figure 2 shows that more than 75% of scans on baselines in the 2017 array that include ALMA had $T_{coh}$ greater than 10 s.

### 2.3. Fourier Coverage

An interferometer like the EHT samples the Fourier transform of the image on the sky. By correlating the data obtained from $N$ stations, $N(N-1)/2$ spatial frequencies are measured.

---

[137] The SEFD of a radio telescope is the total system noise represented in units of equivalent incident flux density above the atmosphere (see Paper III, Equation (3)).

[138] The visibility is the complex two-point correlation of the electric field measured on a VLBI baseline (see Thompson et al. 2017, Chapter 1).

**Table 3**
EHT Station Sensitivities

| Facility | Effective Diameter (m) | Aperture Efficiency ($\eta_A$) | Median Effective System Temperature (K)[a] | Median SEFD$_{1.3}$ (Jy)[a] | $\sigma_{rms-ALMA}$ (mJy)[b] |
|---|---|---|---|---|---|
| Facilities that Participated in the 2017 Observations, Showing Representative Values | | | | | |
| ALMA37[c] | 73 | 0.68 | 76 | 74 | ... |
| APEX | 12 | 0.61 | 118 | 4700 | 2.4 |
| JCMT | 15 | 0.52 | 345 | 10,500 | 3.5 |
| LMT[d] | 32.5 | 0.28 | 371 | 4500 | 2.3 |
| PV 30 m | 30 | 0.47 | 226 | 1900 | 1.5 |
| SMA6[c] | 14.7 | 0.75 | 285 | 6200 | 2.7 |
| SMT | 10 | 0.60 | 291 | 17,100 | 4.5 |
| SPT[d] | 6 | 0.60 | 118 | 19,300 | 4.8 |
| Facilities Joining EHT Observations in 2018 and Later, Using Estimated Values | | | | | |
| GLT | 12 | 0.58 | 120 | 5000 | 2.4 |
| NOEMA | 52 | 0.50 | 270 | 700 | 0.9 |
| KP 12 m | 12 | 0.59 | 310 | 13,000 | 3.9 |

**Notes.**

[a] SEFD = $2kT^*_{sys}/\eta_A A_{geom}$ with $T^*_{sys}$ the effective system temperature and $A_{geom}$ the geometric collecting area of the telescope. The effective system temperature lists representative values for the 2017 Sgr A* and M87 observations, i.e., for median elevations of the sources and typical atmospheric opacities at each facility. It includes effects from elevation-dependent gains and phasing efficiency relevant for some of the telescopes.

[b] With $\sigma_{rms-ALMA}$ the thermal noise for the station's baseline with ALMA using Equation (1) for a bandwidth of 4 GHz and integration time of 10 s (see Figure 2).

[c] The median number of phased antennas in 2017 for the SMA was 6 × 6 m antennas and for ALMA was 37 × 12 m antennas. If all ALMA antennas are phased that would be equivalent to a single 88 m antenna.

[d] The LMT was upgraded from 32.5 m to a full 50 m diameter telescope in early 2018. At SPT the aperture was under-illuminated in 2017 with an effective diameter of 6 m.

As the Earth rotates, those spatial frequencies form tracks in the Fourier plane (i.e., the $(u, v)$ plane) to produce a sparsely sampled Fourier transform of the sky image. With the sensitivity requirements established, one can estimate the baseline coverage that results after the two-stage detection process above (Section 2.2) is followed. Figure 3 shows that the measured spatial frequencies that satisfy the EHT signal-to-noise specification when ALMA is included in the array result in near-full coverage on all baselines in the array, thereby maximizing imaging potential. At the outset of EHT build-out, this full coverage goal served as a design goal based on imaging simulations of synthetic data (see Section 2.1). Subsequent observations in 2017 April confirm that this $(u, v)$ coverage is sufficient to image horizon-scale features (Paper IV).

### 2.4. Time Resolution

Characteristic timescales that affect the evolution of structures in horizon-scale images include the light-crossing time ($T_{light} = GM/c^3$) and the period of the ISCO ($P_{ISCO}$), both of which scale with mass. With an assumed mass of $4.1 \times 10^6 M_\odot$ for Sgr A*, $T_{light} = 20$ s and $P_{ISCO}$ ranges from 4 minutes for a prograde orbit around a maximally spinning black hole to 54 minutes for a retrograde orbit around a maximally spinning black hole. If a mass of $6.2 \times 10^9 M_\odot$ is assumed for M87, $T_{light} = 8.5$ h and $P_{ISCO}$ ranges from 4.5 to 58 days for these same orbits. All of these timescales, while important for imaging and modeling time-variable structures, far exceed the coherence times of the atmosphere, which set the integration intervals and sensitivity requirements for the EHT. Hence, the EHT must already track VLBI observables on considerably finer timescales than $T_{light}$ or $P_{ISCO}$ for the primary cosmic targets, M87 and Sgr A*.

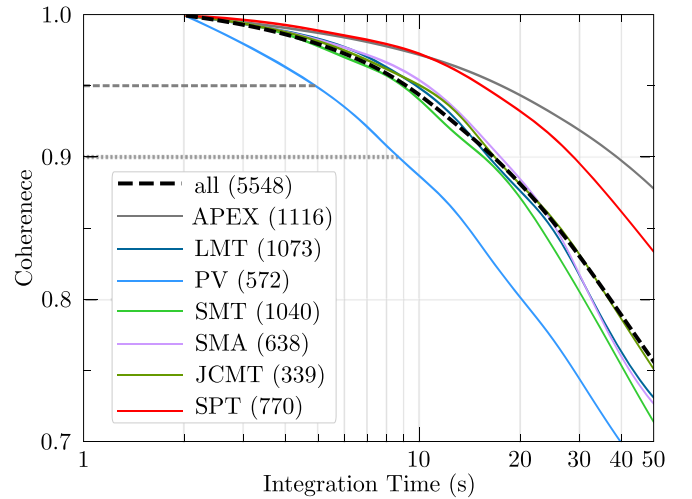

**Figure 2.** Atmospheric coherence time for EHT ALMA baselines. VLBI observations consist of scans, each ~3–7 minutes long (see Figure 12). Each scan in the 2017 EHT observation campaign was divided into sequential coherent integration intervals, which were then incoherently averaged to obtain an estimate of VLBI signal amplitude per scan. As the duration of the coherent interval decreases, the loss in amplitude due to atmospheric phase fluctuations decreases. Solid lines represent the 25th percentile of the coherence distribution for each baseline: 75% of the data exhibit higher time coherence than indicated by the line. The heavy black dashed line shows the 25th percentile of the coherence distribution for all baselines. The light dashed and dotted lines show the 95 % and 90 % coherence limits for the baseline with lowest coherence, which is from PV to ALMA (largely due to low-elevation PV scans on Sgr A* during the observations). The number of scans used to compute the coherence curves are shown next to each station in parenthesis.

### 2.5. Frequency Configuration

The instrumentation at EHT sites is designed to match ALMA's bandwidth and intermediate frequency (IF) ranges.

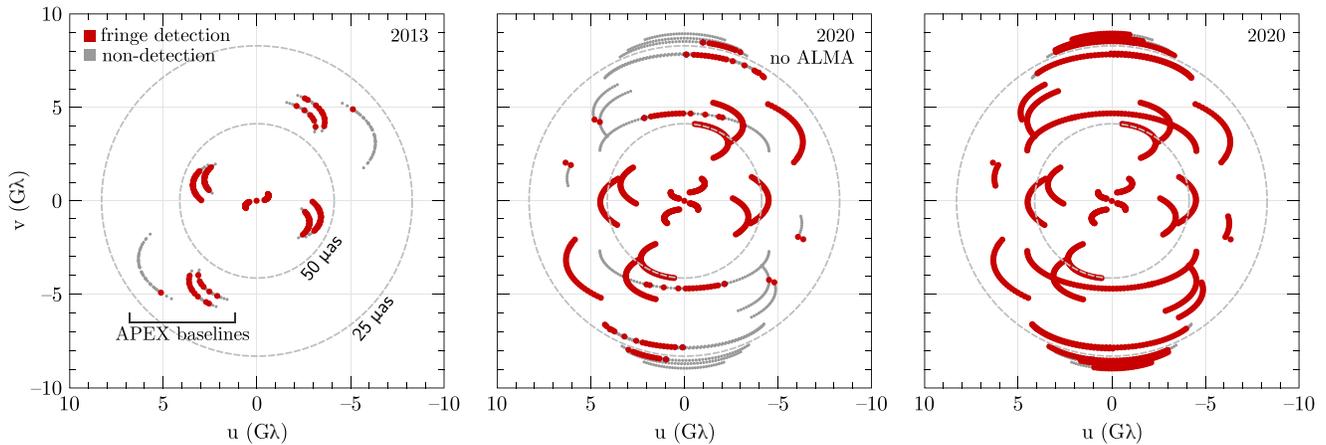

**Figure 3.** Expected EHT Fourier space coverage on Sgr A*. The left panel shows both detections (red) and non-detections (gray) of Sgr A* in the 2013 EHT campaign. Participating telescopes were: APEX, CARMA, JCMT, SMA, and SMT. The dashed circles mark baselines with a fringe spacing equal to 50 $\mu$as (approximately the diameter of the shadow of the SMBH candidate Sgr A*) and 25 $\mu$as. The two remaining panels show simulated EHT observations in 2020: (1) without ALMA and (2) with ALMA. Specifications to determine baseline detections shown are those detailed in Section 2.2. These figures emphasize the benefit of including ALMA in the array: its high sensitivity allows detections for SgrA* on all observed baselines. Because each EHT site requires at least one strong baseline to identify an interferometric fringe and to correct for residual delays, rates, and phase wander, ALMA significantly extends the Fourier coverage and sensitivity even for non-ALMA baselines. Coverage shown in the two right panels corresponds to an array including Chile (APEX, ALMA), Mexico (LMT), France (NOEMA), Spain (PV), Hawaii (SMA, JCMT), Arizona (SMT), and South Pole (SPT). The corresponding baseline coverage of the 2017 observations is shown in Figure 11.

**Table 4**
EHT Frequency Configurations

|  | 230 GHz Band 1SB | 230 GHz Band 2SB | 345 GHz Band 2SB |
|---|---|---|---|
| Nominal Wavelength | 1.3 mm | 1.3 mm | 0.87 mm |
| Lower-sideband Sky Freq. Range (GHz) | unused | 212.1–216.1 | 334.6–338.6 |
| Upper-sideband Sky Freqs. Range (GHz) | 226.1–230.1 | 226.1–230.1 | 346.6–350.6 |
| Local Oscillator (GHz) | 221.1 | 221.1 | 342.6 |
| Intermediate Frequency Range (GHz) | station dependent | 5–9 | 4–8 |
| Recording Rate ( Gbps) | 32 | 64 | 64 |
| Year of First Use | 2017 | 2018 | >2020 |

ALMA antennas are equipped with dual-polarization, sideband-separating receivers with four IF outputs (Section 3.1). The total bandwidth that can be transported back from each ALMA antenna is 16 GHz, thus resulting in a maximum IF bandwidth of 4 GHz per output. The sky frequencies for both the lower- and upper-sideband must overlap perfectly from station to station, which fixes both the local oscillator (LO) frequency and IF frequency ranges at each observatory (Tilanus et al. 2013). Table 4 shows the frequency configurations of the EHT in both atmospheric transmission windows around 230 and 345 GHz.

The main considerations for selecting the chosen LO frequencies are (Marrone et al. 2014) as follows.

1. The tuning range of the receivers at participating facilities.
2. Atmospheric transmission.
3. The avoidance of Galactic $^{12}$CO: $\nu \sim 230.3$–230.8 GHz and $\nu \sim 345.4$–346.1 GHz within the VLBI observing bands.
4. To a lesser degree: avoidance of Galactic $^{13}$CO: $\nu \sim 220.2$–220.6 GHz and $\nu \sim 330.3$–330.9 GHz within the VLBI observing bands.
5. Access to $^{12}$CO spectral lines in an extended tuning range of the receivers above 9 GHz when observing in the 1.3 mm band, or below 4 GHz when observing in the 0.87 mm band.

6. Access to maser lines within the VLBI band, e.g., the SiO maser line at 215.596 GHz ($\nu = 1$, $J = 5 \to 4$).
7. Performance of existing quarter-wave plates used to observe circular polarization.

For ALMA's 230 GHz band, the IF band previously was restricted to a lower limit of 5 GHz (now 4.5 GHz), which resulted in a common IF range across the telescopes of 5–9 GHz, whereas the common IF range for the 345 GHz band is 4–8 GHz.

### 3. Instrumentation

A schematic of the EHT's VLBI signal chain at single-dish telescopes is shown in Figure 4. The front end is typically a dual-polarization sideband-separating receiver in the 1.3 mm or 0.87 mm bands, often the product of a joint development project between the EHT and the telescope facility. These efforts are described in the Appendix. A hydrogen maser provides a frequency reference standard of sufficient stability for mm-VLBI (Section 3.2), and is used to phase lock all analog systems as well as digital sampling clocks throughout the signal chain.

Dedicated block downconverters (BDCs; Section 3.3) mix IF bands coming from the receivers to baseband. Each 4 GHz wide IF band is split and downconverted into two 2 GHz wide sections at baseband. High-bandwidth digital backends (DBEs; Section 3.4) are used to sample two 2 GHz baseband signals each. The sampled data are put into the VLBI Data Interchange Format (VDIF; Whitney et al. 2009) and transmitted to a Mark 6 recorder (Section 3.5) via two 10 Gbps Ethernet (GbE) small form-factor pluggable (SFP+) interfaces.[139] The 16 Gbps output from each DBE matches the recording rate of a single Mark 6 VLBI recorder that records in parallel to 32 hard drives (Whitney et al. 2013). Four DBE—Mark 6 systems are used to sample and record the four 4 GHz wide IF bands coming from the two receiver sidebands and dual polarizations for an aggregate data rate of 64 Gbps.

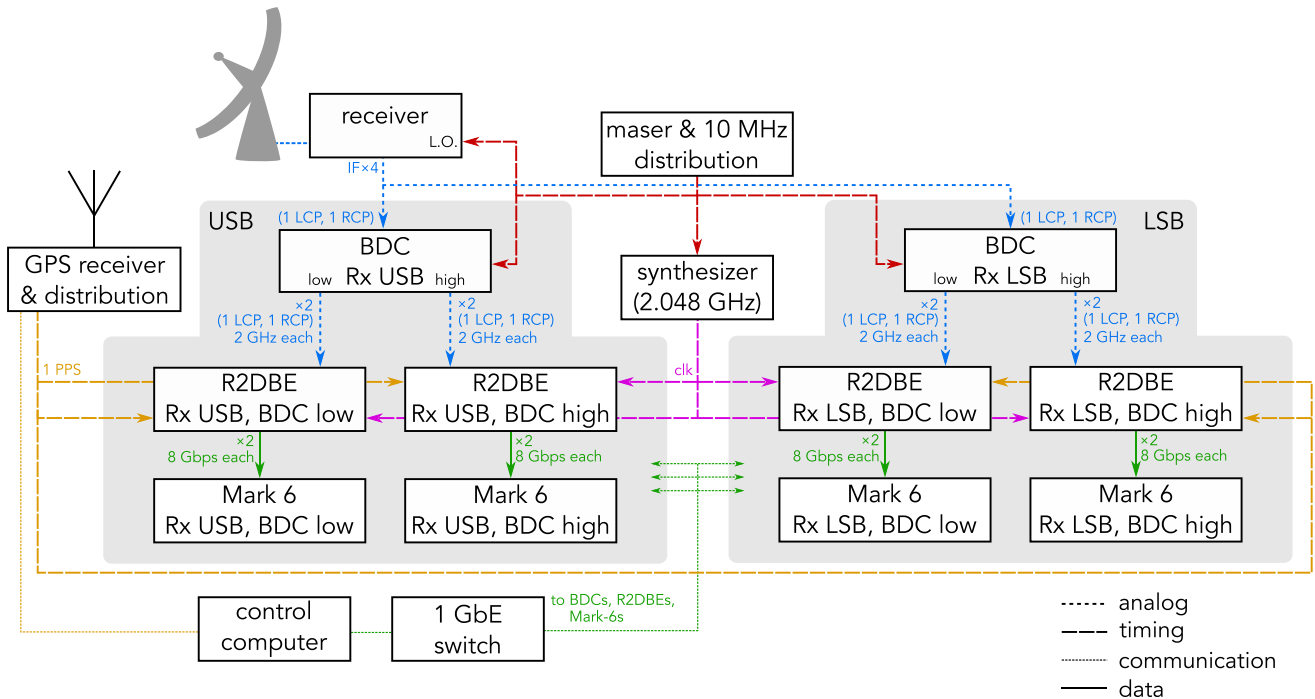

**Figure 4.** System diagram at a single-dish EHT site with 64 Gbps capability. Dual-polarization (left and right circular polarization (LCP and RCP)) and upper- and lower-sideband (USB and LSB) analog IF signals are sent from the receiver. The receiver local oscillator is locked to the station hydrogen maser 10 MHz reference. Block downconverters (BDCs) mix these signals to baseband. R2DBEs sample the analog signals and distribute packetized data to Mark 6 recorders over a 10 GbE network. All timing is locked to a 10 MHz maser reference and synchronized with a pulse-per-second (PPS) Global Positioning System (GPS) signal. The components are controlled through a 1 GbE network. See Figure 8 for a photograph.

For operations at high-altitude, helium-filled hermetically sealed hard drives are used, avoiding the need to build pressure chambers around the recorders. The hard drives collectively accumulated over 10,000 hr of recording at up to 5100 m altitude without a disk failure during the 2017 observations. The full 64 Gbps signal chain thus uses 128 hard drives in parallel, each with a capacity of 6–10 TB for a total storage capacity of about 1 PB. For the stations on the periphery of the network that cannot participate in all scans, 1 PB supports a typical VLBI observing campaign of about 6 days, while stations in the center have a double set of modules, or about 2 PB. Across the array, an EHT observing campaign involving all stations in Table 3 would require a total of about 15 PB of data storage.

Connected-element (sub)millimeter interferometers have more complex systems that use phased-array processors to sum signals from multiple telescopes into a single signal with much greater sensitivity. These processors are integrated with their digital correlators and output the summed signal as VDIF data packets to Mark 6 recorders as at single-dish stations.

The following sections describe each subsystem in more detail.

### 3.1. Receivers

The past three decades have seen the development and widespread use of heterodyne receivers in the millimeter and submillimeter bands based on superconductor–insulator–superconductor (SIS) junctions (e.g., Phillips et al. 1981; Maier 2009; Carter et al. 2012; Tong et al. 2013; Kerr et al. 2014). Over this period, instantaneous bandwidths increased by more than a factor of 30, while noise temperatures decreased by an order of magnitude. Improvements in receiver and antenna reflector technology have combined with the increased recording rates to lay the foundations for a millimeter wavelength VLBI array that is capable of observing targets with a flux density below 1 Jy. In the receivers, high-frequency radiation from the sky is mixed with a pure tone (an LO signal derived from a high-stability frequency reference) and down-converted to an IF using photon-assisted tunneling to transport current across the SIS junction's energy gap. Presently installed SIS-based receivers at observatories typically support IF bandwidths of 4–12 GHz, with higher bandwidth systems becoming available at selected sites.

Heterodyne receivers have two observing bands with sky frequencies centered at $\nu_{LO} - \nu_{IF}$ (LSB) and $\nu_{LO} + \nu_{IF}$ (USB), respectively, from which the signals are folded on top of one another (double-sideband (DSB) receiver). With the use of an RF 90° hybrid, two mixers, and an IF 90° hybrid, modern millimeter receivers can separate the sidebands into two distinct IF channels (sideband-separating (2SB) receiver). The SIS mixers employed at these wavelengths use single-polarization rectangular waveguides that couple to a single polarization of the incident radiation through planar antennas mounted in rectangular waveguides. Dual-polarization receivers employ two independent orthogonal signal chains. Polarizations are split using a wire grid or a waveguide-based orthomode transducer (OMT). The latter can be inserted in front of the mixer blocks and cryogenically cooled, resulting in reduced ohmic losses. Some of the EHT stations use OMTs with circular polarizers, but the majority use quarter-wave plates in front of the receivers. ALMA records in dual orthogonal linear polarization, which is converted to circular polarization in post-processing as described in Section 5.

### 3.2. Hydrogen Maser Frequency Standards

As with connected-element arrays, VLBI relies on the fundamental ability of radio band receivers to accurately preserve the phase of detected cosmic radiation. For connected-element arrays, a common LO, derived from a station frequency reference, is used for all antenna receivers. On long VLBI baselines, sharing a common frequency reference is currently not technically feasible, and each VLBI site must use its own clock. This practice imposes a strict requirement on the stability of the frequency standards used for VLBI. Generally, one requires $\omega \tau \sigma_y(\tau) \ll 1$, where $\omega$ is the observing frequency in rad s$^{-1}$, $\sigma_y(\tau)$ is the square root of the Allan variance (the Allan standard deviation), and $\tau$ is the integration time. This limit keeps rms phase fluctuations well below 1 rad (Rogers & Moran 1981).

Because of their excellent stability on timescales that match VLBI integrations (about 10 s, Figure 5), hydrogen masers are used almost without exception as VLBI frequency references worldwide. At $\sigma_y(10\,\mathrm{s}) = 1.5 \times 10^{-14}$, VLBI observations at wavelengths down to 0.87 mm can be carried out with coherence losses (Loss $\simeq 1 - e^{-(\omega \tau \sigma_y(\tau))^2}$) limited to <5 % on a baseline where stations at either end are equipped with similarly stable masers (Doeleman et al. 2011). The use of hydrogen masers ensures that phase fluctuations in the VLBI signal due to atmospheric turbulence are typically the dominant source of coherence loss in EHT observations (Rogers et al. 1984; see also Figure 2).

The hydrogen masers used in the EHT are monitored for stability using two tests. The first compares the stability of the 10 MHz frequency reference output of the maser with that of a high-precision quartz oscillator, known to have a similar performance as a maser on 1 s timescales. This in situ measurement confirms nominal maser operation at 1 s integration times, and extrapolation then yields an estimate of maser performance at 10 s (see Figure 5). A second test is to routinely monitor the offset between 1 PPS signals derived from the maser and a GPS receiver. GPS stability on 1 s timescales is poor due to variations induced by the ionosphere, but on longer timescales (>10$^4$ s) GPS precision, which is referenced to the average of many atomic clocks, will exceed that of any single maser at an EHT location. An observed linear drift (Figure 5) over month to year timescales confirms long-term maser stability and also provides an instrumental delay-rate that is required for later interferometric detection (see Section 5).

### 3.3. Block Downconverters

A BDC design was developed to mix selected pieces of the incoming IF band from the receiver (in the range 4–9 GHz) down to baseband (0–2 GHz) for sampling. For observing at

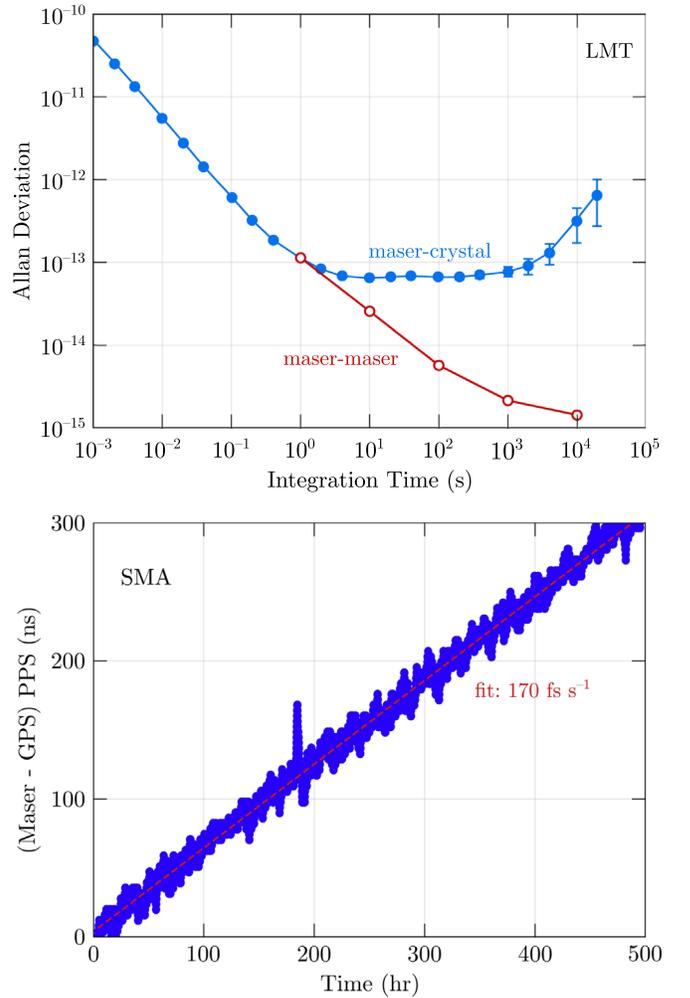

**Figure 5.** Top panel: measured Allan standard deviation, $\sigma_y(\tau)$, of the hydrogen maser at the LMT compared to a precision quartz oscillator. The open red points are the manufacturer specifications of the hydrogen maser when compared to another maser. At a 1 s integration time, the quartz oscillator and maser have similar stability, so the measurements indicate that the maser is meeting its specifications as installed at a coherence time of ~1 s. The flattening of the $\sigma_y(\tau)$ curve beyond 10 s is due to the decreased stability of the quartz crystal. The extrapolation of Allan deviation from short integration times to 10 s indicates that the maser meets the stability goal of $\sigma_y(10\,\mathrm{s}) = 1.5 \times 10^{-14}$. Bottom panel: the long-term drift of the maser at the SMA compared to GPS, measured by differencing the 1 PPS ticks from the maser and local GPS receiver. The vertical width of the trace is due to variable ionospheric and tropospheric delays of the GPS signal (including the excursion near hour 200), while the long-term trend represents the frequency error of the maser. The drift measured from this plot, and its effects on the fringe visibility, are removed during VLBI correlation.

1.3 mm, bandpass filters select 5–7 GHz and 7–9 GHz bands that are mixed against a LO at 7 GHz to convert to baseband. Similarly, for observing at 0.87 mm, 4–6 GHz and 6–8 GHz bands are mixed against an LO at 6 GHz to convert to baseband. The LO is phase locked to a 10 MHz reference from the station frequency standard. Coaxial relays in the output stage select which set of baseband outputs is connected to the data acquisition system, and the LO of the unused set is muted to avoid interference. The downconverter has two duplicate chains of filters and mixers for processing two of the four receiver IF channels, and two such downconverters are used in a complete system. A simplified schematic of a BDC is shown in Figure 6.

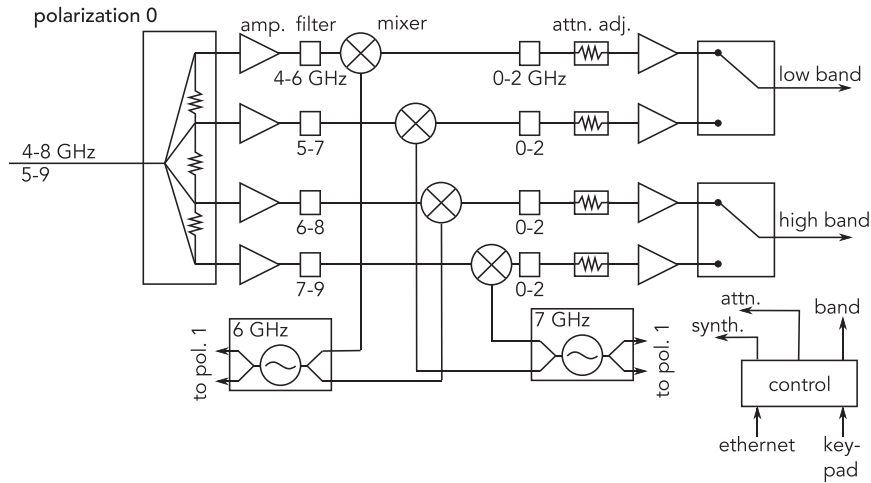

**Figure 6.** Simplified schematic diagram of the BDC. Only one of the two (identical) polarization channels is shown, and several intermediate amplification stages are omitted. Two local oscillators maser-locked and tuned to 6 GHz and 7 GHz are used to mix all the filtered IF bands between 4 and 9 GHz to baseband (0–2 GHz). For 230 GHz operation, a 5–9 GHz IF band is split with output of 5–7 GHz ("low-band") and 7–9 GHz ("high-band"). For 345 GHz operation, a 4–8 GHz IF band is split with an output of 4–6 GHz ("low-band") and 6–8 GHz ("high-band"). The 6 and 7 GHz local oscillator signals are also sent to the other polarization channel. The attenuation and band selection can be set using a control panel or remotely via an Ethernet connection.

The maximum conversion gain of each channel is about 23 dB, and it can be adjusted downward to about −8 dB in 0.5 dB steps to match the output power of the BDC with the input dynamic range of the digitizers. At a nominal output power of −7 dBm, the BDC operates in the linear regime over the entire programmable range of the attenuators. An 8-bit controller provides control of the synthesizers, digital attenuators, and coaxial relays, and interfaces to a keypad and display unit for manual operation, as well as to an Ethernet port for remote control.

### 3.4. Wideband VLBI Digital Backend

Coherent received station data in modern VLBI systems are recorded digitally. The VLBI instrument that digitizes and formats the analog received signal for recording is termed the digital backend (DBE). The high bandwidths that enhance the EHT sensitivity require proportionately fast digital sampling speeds. The timing of the samples is an implicit time stamping of the data, so the sampler clock must be timed with maser stability and precision.

For several generations of instrumentation the EHT has based its digital hardware on open-source technology shared by the Collaboration for Astronomy Signal Processing and Electronics Research (CASPER;[140] Hickish et al. 2016). The open-source hardware currently in use includes five gigasample-per-second (Gsps) analog-to-digital converter (ADC) boards, based on an integrated circuit that interleaves four cores, each with a maximum sample rate of 1.25 Gsps. The ADC circuit design and hardware testing is documented in Jiang et al. (2014). The compute hardware platform is the CASPER second-generation Reconfigurable Open Access Computing Hardware (ROACH2). The ROACH2 uses an FPGA as its digital signal processing engine.[141]

The current EHT DBE instrument was developed in 2014, and is called the ROACH2 Digital Back End (R2DBE;

Vertatschitsch et al. 2015). The R2DBE samples two analog channels at 4.096 Gsps, each with 8-bit resolution, i.e., two 2.048 GHz wide bands, before re-quantizing the samples with 2-bit resolution, packing the data in VDIF format, and then transmitting it over two 10 GbE connections to the Mark 6 recorder.[142] Each R2DBE transmits two 8 Gbps data streams to a Mark 6 for recording. Accurate timing and synchronization is achieved by referencing the ROACH2 clocks to the maser via an external EHT-developed 2.048 GHz synthesizer.

The per-channel signal processing flow within the R2DBE is shown in Figure 7. Gain, offset, and sampler clock phase mismatches between the cores of the ADC produce spurious artifacts in the spectrum of the digitized signal, most prominently a spur at one quarter of the sample rate frequency that is caused by interleaving imperfection of the quad-core ADC. A calibration routine is performed at the start of each observation to tune the distribution of 8-bit samples for each core in offset, gain, and phase, removing these artifacts (Patel et al. 2014). A digital power meter is implemented in the firmware to calculate input power every millisecond, which is useful for pointing and calibration measurements.

An important design consideration is the optimal number of bits per sample recorded onto storage media at the telescopes. This is determined by trading off the sensitivity increase realized through sampling more bandwidth at higher precision (see Equation (1)) with the cost of media required to store the data (for a detailed explanation, see Thompson et al. 2017). For the EHT, a maximum aggregate bandwidth of 16 GHz is set by the characteristics of ALMA receivers. When bandwidth is limited in this way, one examines the increase in sensitivity per additional unit media. For 1-bit recording (two-level sampling), the digital efficiency ($\eta_Q$ in Equation (1)) is 0.64. Moving to 2-bit recording (four-level sampling) increases $\eta_Q$ to 0.88 at the expense of doubling the required recording media, for a 38 % sensitivity increase per additional unit media—this is approximately equivalent to doubling the front-end bandwidth and keeping 1-bit sampling. Tripling the media cost and recording 3 bits (eight-level sampling) delivers only a 25.5 % increase in sensitivity per additional unit media. The 2-bit sampling scheme was chosen as it ensures ample margin for sensitivity requirements while minimizing recording media, which dominates the cost of VLBI correlation (Deller et al. 2011).

---

[140] For more information on CASPER, please see https://casper.berkeley.edu/.

[141] See https://www.xilinx.com/support/documentation/data_sheets/ds150.pdf.

[142] An alternative DBE that is planned to be deployed at some EHT sites is the third-generation Digital BaseBand Converter (DBBC3; Tuccari et al. 2014).

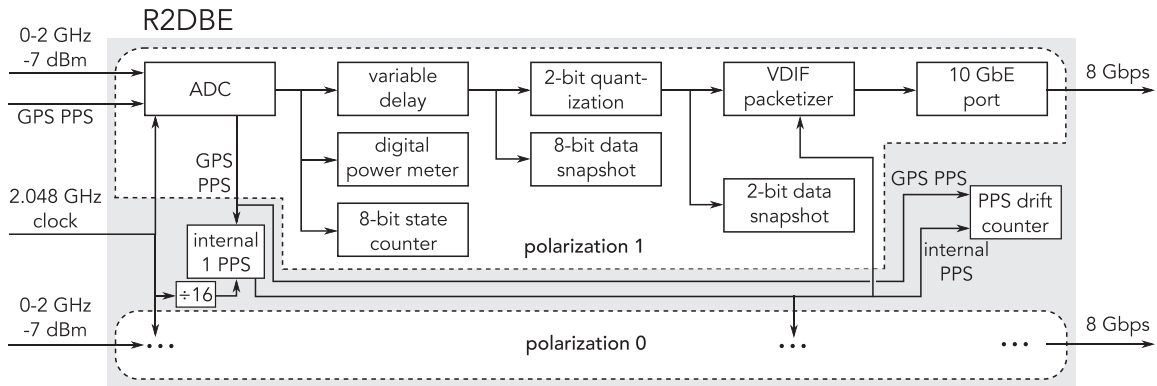

**Figure 7.** Functional diagram of the R2DBE. A pair of ADCs sample two channels of 2.048 GHz Nyquist bandwidth IF bands, typically representing two antenna polarizations. After requantization of the samples to 2 bits, the data are distributed in VDIF packets over a 10 GbE network to Mark 6 recorders. The maximum FPGA clock speed is too slow to process the ADC output stream rate of 4096 megasamples per second in series, so 16 ADC samples are transferred from the sampler in parallel on each FPGA clock cycle and are processed in parallel though the FPGA, which is clocked at a sixteenth of the sample clock, or 256 MHz. The FPGA and sampler clocks are locked to the maser reference frequency. In addition to the sample clock and dual IF inputs, each ADC circuit board is also equipped with a low-frequency synchronization input. On one of the ADC boards, the synchronization input is connected to GPS PPS. An internal PPS for VDIF time stamping is synthesized from the 2.048 GHz maser-referenced clock and synchronized to the GPS real-time PPS time-tick once at the beginning of an observation. The internal PPS drift relative to GPS real-time PPS (Figure 5) is measured by a counter and is available to be read from a register over Ethernet, as well as copied into the VDIF data-header, as it is a prior parameter for VLBI correlation.

In a 2-bit system, the noise voltage thresholds of the four sampling levels should be properly set to maximize digital processing efficiency. A proper setting ensures that all four levels are optimally populated for a given input voltage signal from the telescope receiver system (Cooper 1970; Thompson et al. 2017). The processing efficiency is not a sensitive function of these thresholds, but the statistics of the level populations can vary significantly as elevation and weather condition changes cause fluctuations in the receiver power arriving at the sampler. For this reason, a calibration of the sampler threshold setting are performed periodically during calibration scans to maintain optimal settings.

### 3.5. High-speed Data Recorders

The Mark 6 VLBI Data System (Whitney et al. 2013) is a packet recording system used to store the R2DBE output streams to hard drives. The recorder captures the two 8 Gbps streams from the R2DBE using commercial 10 Gbps Ethernet network interface cards. It strips the internet packet headers and stores the payload containing VDIF data frames. For sustained recording at 16 Gbps, each Mark 6 recorder writes the data in time slices across 32 hard drives with a round-robin algorithm. The disks are mounted in groups of eight in four removable modules for ease of handling and shipping. For the EHT, four such recorders are configured in parallel to achieve an aggregate data capture rate of 64 Gbps.

Mark 6 recorders and modules are commercially available and in use in other VLBI applications. Several modifications were required for EHT use at high altitude, as hardware specifications typically state 10,000 ft as the maximum operating altitude. The low ambient air density necessitates sealed, helium-filled hard drives, both for the system disk for recorders and also in all modules onto which data are recorded. In addition, the Mark 6 interior was modified to direct a high-volume airflow onto the CPU, network, and data interface cards, which were found to be sensitive to overheating at altitude.

A photograph of the EHT VLBI backend (BDCs, R2DBEs, and Mark 6s) at the PV 30 m telescope is shown in Figure 8. Recorders used for earlier EHT campaigns included the Mark 5C system, which was capable of capturing only a 4 Gbps data stream onto two modules. The bandwidth of the EHT backend since 2004 is plotted in Figure 9 and represents a 64-fold increase, corresponding to an eight-fold improvement in sensitivity.

### 3.6. Phased Arrays

To use the total collecting area available at connected-element interferometers that participate as single stations in the EHT VLBI network (ALMA, SMA, and, in the future, the Northern Extended Millimeter Array (NOEMA)) these arrays coherently add the signal received from the target source by each antenna and record as if from a single antenna. Practical constraints on the maximum data rate that can be recorded at each site require that this summation be performed in real time.

Forming a coherent sum requires correcting for deterministic delays, such as geometric and known instrumental delays, as well as non-deterministic delays, which at EHT frequencies are significantly affected by the distribution of water vapor in the atmosphere. The atmospheric delay varies over time, antenna location, and direction so that accurate compensation can only be achieved through the use of in situ calibration methods.

Phasing systems were developed for the SMA and ALMA (see Appendix), and these observatories participated as phased arrays since 2011 and 2017, respectively. A phasing system for NOEMA is in the process of being implemented and commissioned for VLBI. Figure 10 shows the SMA phasing efficiency, i.e., the fraction of beamformed power compared to ideal phasing, estimated using Equation (2), achieved over the course of several scans during one night of the 2017 EHT campaign. For most of the scans, the efficiency is well above 0.9 (see inset). Typical phasing performance of the ALMA array can be found in Matthews et al. (2018).

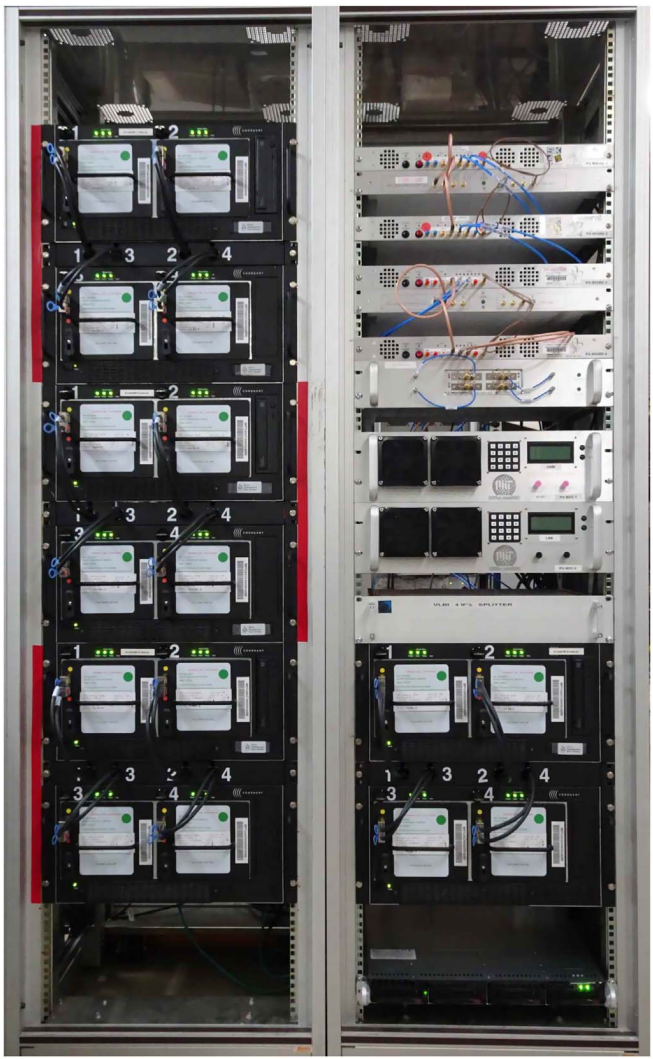

**Figure 8.** EHT digital VLBI backend as installed at the Institut de Radioastronomie Millimétrique (IRAM) PV 30 m telescope in Spain. The upper portion of the right-hand side rack holds the four R2DBE units. Two block downconverters are installed near the middle. The VLBI backend computer is mounted on the bottom right. The rack on the left and the lower portion of the rack on the right hold the four Mark 6 recorders with four disk modules each, providing a total of about 1 PB in data storage.

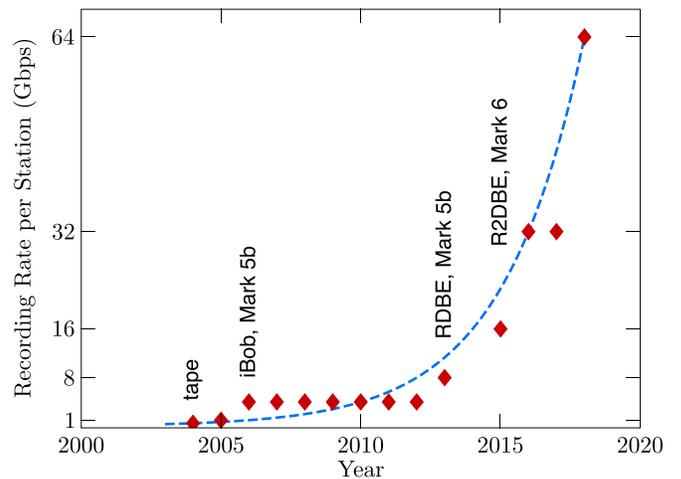

**Figure 9.** Recording rate of EHT observations over time. As of 2018, EHT stations record at 64 Gbps, equivalent to a doubling of recorded bandwidth every two years for over a decade (blue curve). The high bandwidth, a result of linking EHT instrumentation to industry trends and commodity electronics, is a crucial component of the EHT's sensitivity, enabling detections on long baselines, providing resilience of the network against poor weather and low-elevation targets, and allowing detections from all stations to ALMA within short atmospheric coherence times (see Section 2.2).

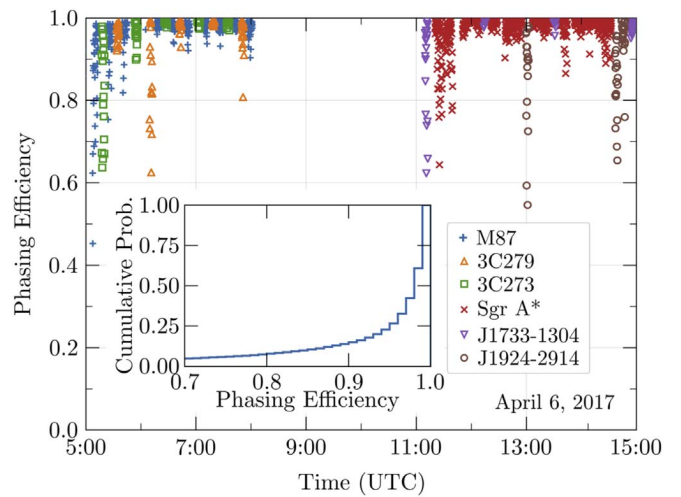

**Figure 10.** Phasing efficiency of the SMA on various sources over the course of one observing schedule. The inset shows the cumulative distribution of phasing efficiency of the SMA over the entire 2017 EHT observing campaign.

## 3.7. Setup and Verification

As part of the integration of the new VLBI systems in 2015, most station backends were configured at Haystack Observatory prior to shipment to the telescopes. New equipment still typically passes through Haystack for initial inspection, checkout, and configuration. Each site takes responsibility for the installation of the equipment and ensuring that the connections to facility hardware work to specification and are not affected by factors such as telescope motion.

Before each observing campaign, each site goes through a comprehensive setup and verification procedure, which includes the completion of a checklist by site operational staff. The procedure verifies operation of the BDC, R2DBE, and Mark 6 systems, and the exact frequency and lock status of LOs, which set the exact observed sky band, and the ADC clock that time stamps the data. The check also includes the coherence and drift of the hydrogen maser time standard as this clock rate measurement is needed during correlation for fringe-stopping.

Due to the remote nature of many of its sites and the large data volume recorded, the EHT lacks real-time verification of fringes. Shipping data from remote stations to the central correlator and processing takes several days at minimum, and many months in the worst case (from the South Pole; see Appendix A.11). If a key system fails, it is possible for a site to take data that never result in fringes, making careful testing and retesting of subsystems throughout the observation absolutely crucial. Data from brief observations (10–30 s) can be transferred from sites with fast internet connections for near-real-time fringe verification. While possible, this requires robust data transfer connections.

The most complete in situ full-system check consists of the injection of a test tone at a known frequency in front of the receiver. A short data segment is recorded, and the autocorrelations and zero-baseline cross-correlations are inspected to verify that the test tone appears at the correct frequency and with the correct profile. Further, the tone in the baseband is mixed down to 10 kHz using a third LO and compared to a 10 kHz tone derived from the maser 10 MHz reference. The two should be phase locked when examined on an oscilloscope, which verifies the phase stability of the whole system. This test is exquisitely sensitive to small disturbances in the system. At the JCMT an equivalent test is done by verification of local connected-element fringes with the SMA.

Results and logs from the setup and verification are centrally archived and available to the correlator centers for the interpretation of station issues and for fault diagnosis when data quality issues emerge.

### 3.8. The Array

The EHT included eight observing facilities in 2017. Three additional facilities have since joined (GLT) or will soon join (KP 12 m and NOEMA), and two facilities (CARMA and the CSO) have participated in past EHT observations but have now been decommissioned. Properties of these facilities are summarized in Tables 2 and 3, and the Appendix. The baseline Fourier coverage provided by the array in 2017 for M87 is shown in Figure 11. Table 5 shows the actual performance of the array on scans of M87 during the 2017 science observations. The scan-averaged thermal noise within one 2 GHz frequency band was significantly better than 1 mJy on most baselines to ALMA, and generally a few mJy on baselines excluding ALMA. These achieved sensitivities, combined with the realized $(u, v)$ coverage in Figure 11, confirm that the EHT has met its essential specification in providing an array that can image features on the scale of an SMBH shadow.

## 4. Observing

EHT science observing campaigns are scheduled for March or April when Sgr A* and M87 are night-time sources and the weather tends to be best on average over all sites. In addition, ALMA tends to be in a more compact configuration with better prospects for including more antennas in a phased array for increased sensitivity. Test and commissioning runs are scheduled a few months prior to campaigns to ensure that VLBI equipment, which may be dormant for months at a time, is operational.

### 4.1. Weather

Weather is an important consideration for VLBI observations at millimeter wavelengths. Most of the EHT observatories are located in the northern hemisphere, and those stations have especially large weather variations between seasons. Opacity and turbulence are typically lowest at night and in winter and early spring. At many sites, and particularly the connected-element arrays, the reduced atmospheric turbulence during night-time hours is essentially required. To protect against inclement weather, the EHT uses flexible observing with windows that are about twice as long as the intended number of observing nights. Within the window, a few hours before the start of observing each night, weather conditions are reviewed at all sites and a decision is made whether or not to observe that night. When an EHT night is not triggered, it is often possible for the time to be used for other observing programs at an observatory.

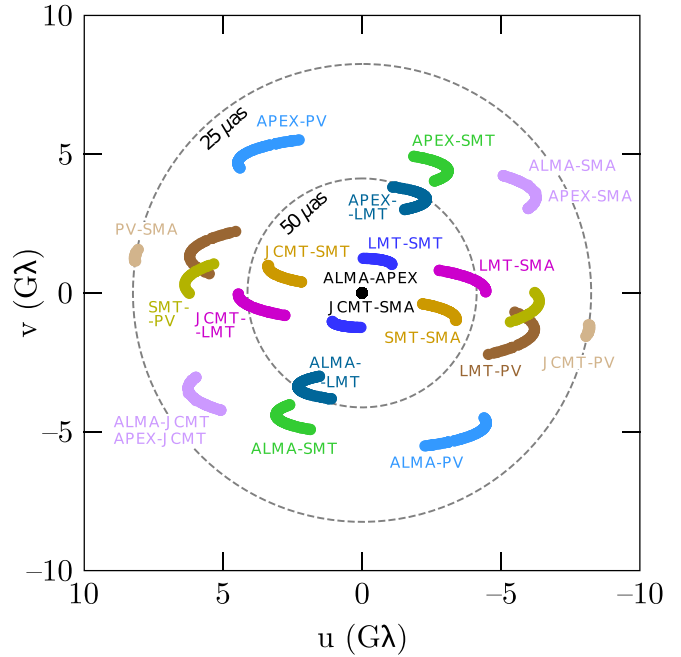

**Figure 11.** Aggregate EHT baseline coverage for M87 over four nights of observing with the 2017 array. Only detections are shown. The dashed circles show baseline lengths corresponding to fringe spacings of 25 and 50 $\mu$as. See Paper III for details.

**Table 5**
Median Thermal Noise (mJy) for Observations in 2017

|  | APEX | JCMT | LMT | PV | SMA6 | SMT | SPT |
|---|---|---|---|---|---|---|---|
| ALMA37 | 0.53 | 0.80 | 0.34 | 0.43 | 0.54 | 0.57 | 1.75 |
| APEX | ⋯ | 9.99 | 4.24 | 4.22 | 6.49 | 6.68 | 14.51 |
| JCMT | ⋯ | ⋯ | 5.52 | 10.24 | 9.60 | 10.17 | 16.14 |
| LMT | ⋯ | ⋯ | ⋯ | 3.22 | 3.63 | 3.91 | 11.57 |
| PV 30 m | ⋯ | ⋯ | ⋯ | ⋯ | 7.98 | 6.11 | 14.53 |
| SMA6 | ⋯ | ⋯ | ⋯ | ⋯ | ⋯ | 6.64 | 15.33 |
| SMT | ⋯ | ⋯ | ⋯ | ⋯ | ⋯ | ⋯ | 17.73 |

**Note.** Median scan-averaged thermal noise per baseline in mJy for all M87 detections in 2017 April. Entries for the SPT reflect the median thermal noise on 3C 279, since the SPT cannot observe M87 due to its geographic location. Scan durations were three to seven minutes. For more details, see Paper III.

### 4.2. Scheduling

The process of scheduling starts with the list of approved target sources, time allocations at EHT telescopes and ALMA, and the dates of the observing window (Section 4.1). Schedule construction has to satisfy multiple requirements:

1. a high total on-source time for each target,
2. a wide range of parallactic angles sampled for polarimetry,
3. long baseline tracks on all sources within the limited number of observing days,
4. randomized scan lengths and scan start times on Sgr A* for periodicity analysis,
5. good baseline coverage in the $(u, v)$ plane, and
6. regular gaps for telescope pointing and calibration.

AGN sources are chosen as calibrators based on their brightness, compactness, and proximity to the target sources. The calibration sources are used for multiple purposes,

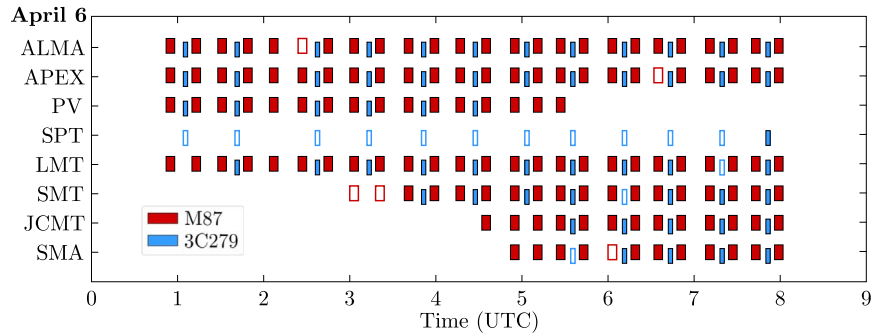

**Figure 12.** EHT 2017 observing schedule for M87 and 3C 279, covering one day of observations (April 6). Empty rectangles represent scans that were scheduled, but were not observed successfully due to weather, insufficient sensitivity, or technical issues. The filled rectangles represent scans corresponding to detections available in the final data set. Scan durations vary between 3 and 7 minutes, as reflected by the width of each rectangle, and scans are separated by periods of time used at each site for local pointing adjustments and calibration measurements. For this schedule, 3C 279 was the calibrator source for the primary target, M87. For information on other observing days, see Paper III.

including fringe finding, bandpass calibration, and polarization calibration.

Observing blocks are created for each of the sources, and these blocks are merged into tracks, each corresponding to a single night of observing. Depending on time allocations and Fourier-coverage needs, multiple blocks on the same source may appear in the same track or in different tracks. Each observatory has different needs for overhead because the EHT is an inhomogeneous array. Overhead accommodates time for pointing, focus, and primary and secondary flux observations at all sites, plus phase-up time and array-polarimetric calibrations at the phased arrays. A typical EHT schedule (Figure 12) records VLBI scans with a duty cycle of approximately 50%, with a substantial fraction of those scans on targets strong enough to use for array calibration.

### 4.3. Monitoring and VLBI Backend Control

The EHT developed a centralized monitoring tool called `VLBImonitor` to visualize observing status, collect ancillary calibration and weather data from each site, and provide real-time and predicted weather information from meteorological services. The information collected, logged, and displayed includes atmospheric and local weather conditions, observatory metadata such as telescope coordinates and on-source status, system temperature measurements, opacity measurements, digital backend and recorder state information, and comments from on-site observers and operators. Communication with the metadata server (by the software clients at each site or via a web interface) uses the JSON-RPC (remote procedure call) protocol over HTTPS. A "masterlist" defines metadata parameters that are accepted (white-listed) by the server and all their properties (e.g., data type, measurement cadence, a function that evaluates if the current value is valid or invalid, and units).

The `VLBImonitor` software[143] was introduced for the 2017 observations. In 2018, a VLBI backend computer and network were added at the sites to provide a common monitoring and control platform at each station. Both monitoring and control remain in active development. At present, the EHT deploys specialist teams at each station, which, in practice, limits the observing window to about 12 days. It is a long-term objective for the EHT to both increase the length of this window and to conduct VLBI observations without the need for on-site specialists, after an initial setup and verification by local and remote experts. A critical component toward this goal is the implementation of comprehensive remote monitoring of the stations and VLBI equipment, as well as remote control of the VLBI backend.

### 5. Correlation and Calibration

The recorded data modules at all sites are separated by frequency band, with the "low-band" shipped to the VLBI correlator at MIT Haystack Observatory in Westford, Massachusetts, USA, and the "high-band" to the correlator at Max-Planck-Institut für Radioastronomie (MPIfR) in Bonn, Germany. Correlation is performed using the Distributed FX (DiFX) software correlator (Deller et al. 2011) running on clusters of more than 1000 compute cores at each site, and is split between the two sites to speed processing and allow cross-checks. At least as many Mark 6 playback units are needed at each correlator as there are stations in the EHT. The Mark 6 playback units at the MIT correlator are connected via 40 Gbps data links. A 100 Gbps network switch then delivers data to the processing nodes using 25 Gbps links. At MPIfR the internode communication, which includes the Mark 6 playback units, is realized via 56 Gbps connections, exceeding the maximum playback rate of the Mark 6 units of 16 Gbps.

Each 2 GHz observing band is correlated independently, with multiple passes required to correlate the full 4 or 8 GHz in an experiment. The correlation coefficients between pairs of antennas are calculated after correcting for an a priori clock model (Earth rotation, instrumental delays, and clock offsets and drift rates). All sites except ALMA record left and right circular polarizations (L/R) producing cross-correlations in the standard circular basis (LL, LR, RL, RR). ALMA antennas, however, are natively dual-linear polarization (X/Y), so a linear-to-circular polarization conversion is performed on ALMA baselines *after* VLBI correlation. `PolConvert` (Martí-Vidal et al. 2016) is a software routine developed for this purpose as part of the program to phase the ALMA array for VLBI operation (Matthews et al. 2018; Goddi et al. 2019). This routine forms linear combinations of the cross-polarization-basis correlator products (XR, YR) and (XL, YL) with ±90° phase shifts introduced to the visibility phase to produce complex visibilities in the circular-polarization basis, for each integration time. The input correlator products to `PolConvert` are equalized in gain and phase, and the X–Y channel phase differences are removed before polarization conversion.

---

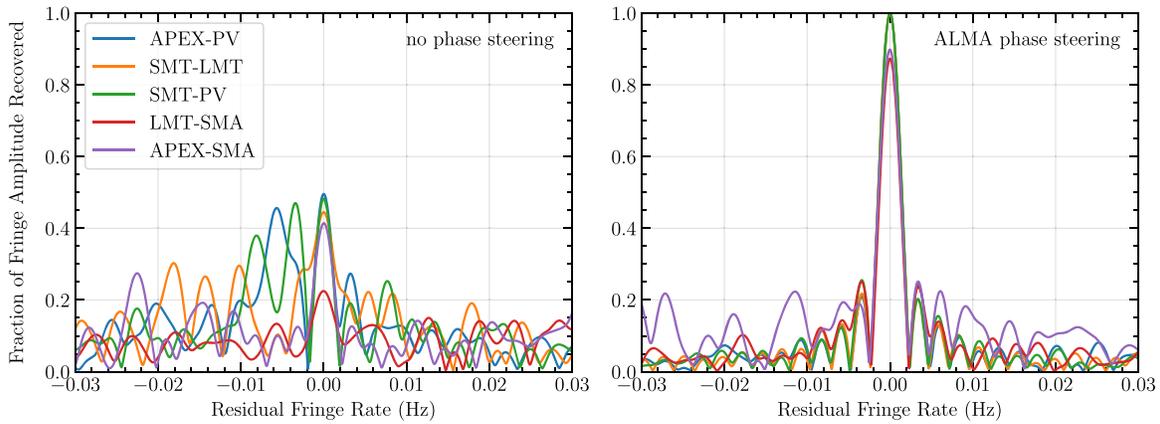

**Figure 13.** Fringes detected on EHT baselines before and after phase steering using the ALMA phased array as a reference station. Shown here are residual fringe-rate spectra, where the peak occurs at a single fringe-rate that is consistent with linear clock drift over the entire scan. Stochastic atmospheric phase variations result in fringe-rate variation over time, spreading the signal in the plot on the left. To recover the total fringe amplitude, baselines to ALMA are used to phase reference the array at a short timescale. Data shown are from a VLBI scan on a quasar calibrator (3C 279) obtained during EHT observations in 2017 April. The data are taken at low elevation, and at several sites during the afternoon and early evening when the atmosphere is often unstable. Baselines to ALMA are able to phase steer the remaining sites in the array at an effective timescale of 1–3 s depending on S/N. At SMA due to the mid-afternoon local observation, phasing is particularly challenging. This leads to an amplitude efficiency of about 90 % on SMA baselines, as well as increased measurement thermal noise for the APEX-SMA spectrum with a full-scan S/N of only ∼10.

This X–Y channel equalization is performed using standard calibration techniques for connected-element ALMA operation (for details see Paper III).

To reduce data volumes and increase S/N, the cross-products computed in the correlator are averaged both in time and frequency. These complex cross-products are then fitted for fringe-delay (linear change in phase versus frequency) and fringe-rate (linear change in phase versus time), which are then removed. Residual fringe-delay and fringe-rate are due to several factors. Largest among these are residual clock drifts, instrumental electronic delays, and atmospheric phase fluctuations that can result in rapidly varying fringe-rate residuals. For the compact, narrow-field EHT continuum targets, delay and rate variability due to intrinsic source structure is expected to produce much smaller residuals. The averaging time and bandwidth in the correlator is thus set to ensure that any coherence losses due to delay or rate variations are negligible, or equivalently that such variations can be tracked both in time and frequency. For EHT observations, the typical averaging bandwidth is 0.5 MHz and the averaging time is 0.4 s, allowing residual delays up to $\pm 1$ $\mu s$ and residual rates up to $\pm 2.5$ Hz. These settings enable the isolation and identification of instrumental spectral features, and also the ability to track atmospheric phase variations. This averaging in both time and frequency, results in a decimation of data volumes by more than a factor of 1 million.

After the initial correlation, the data are further processed through a pipeline that results in final data products for use in imaging, time-domain analyses, and modeling (Blackburn et al. 2019; Janssen et al. 2019). During fringe detection, the excess in correlated signal power due to the source on a VLBI baseline is identified, and the complex Fourier component of source brightness distribution is measured. As outlined in Section 2.2, high signal-to-noise detections on baselines to ALMA can be used to remove atmospheric phase fluctuations on non-ALMA baselines. Phase stabilizing the array in this way allows coherent integration of the VLBI signal on non-ALMA baselines beyond the atmospheric coherence time. Figure 13 demonstrates the process on real EHT data during relatively poor weather conditions, confirming the general approach and basis for the specifications in Section 2. Precise estimates of the observed systematic errors during the 2017 April EHT campaign are detailed in Paper III.

Subsequent calibration steps convert the correlation coefficients derived from fringe detection to correlated flux density (Jy). This is accomplished through use of a priori information about station sensitivities, and by the application of self-calibration techniques to correct for variations in telescope gain over time and frequency. In cases where two EHT telescopes are in close geographical proximity (e.g., ALMA-APEX and SMA-JCMT), additional calibration constraints can be derived from the resulting baseline redundancy and using only general assumptions about the observed source (Paper III).

## 6. Future Developments

The EHT array is continuing to develop. With the ability to image SMBHs on horizon scales now confirmed (Papers I; II; III; V; VI), the focus of EHT development will shift to enabling observations that can refine constraints on fundamental black hole properties, processes of black hole accretion and outflow, and tests of general relativity. This depends on achieving higher angular resolution, enhancing image fidelity, and enabling dynamic imaging of time-variable phenomena. Higher resolution will allow more detailed studies and modeling of sub-horizon structures as well as sensitive tests for asymmetries in shadow features. Greater image fidelity will bring fainter emission near the horizon into focus for the study of accretion and jet processes, and it will enable a sensitive comparison across imaging epochs, which is especially germane for M87 with its dynamical timescale of days to weeks. For Sgr A*, the light-crossing time of ∼20 s requires a dynamic approach to image reconstruction, with the potential of observing the near real-time evolution of a black hole.

The planned and in-progress addition of telescope facilities at new geographic sites will improve the $(u, v)$, or Fourier, coverage, and thus imaging fidelity. Over the course of the next two years, the EHT expects to add two more facilities: a beamformed NOEMA in France, which, once completed, will be the equivalent of an approximately 50 m dish, and a 12 m diameter dish on Kitt Peak in Arizona. A newly realized and important consequence of designing high-bandwidth systems is that adding telescopes with modest diameters ($\simeq$6 m) creates VLBI baselines with sufficient sensitivity to detect the primary EHT targets. An expansion of the EHT can, therefore, include the possibility of deploying numerous smaller apertures, enabling not only improved image quality but also snapshot capability, which is a precursor to constructing black hole movies (Bouman et al. 2018; Johnson et al. 2018). Additional plans to enhance the data throughput of VLBI backend and phased array systems are underway, linked to new generations of wideband millimeter and submillimeter receivers and industry trends that will allow for rapid design and implementation. With the inclusion of more sites and wider bandwidths, the computational requirements for VLBI correlation will increase: linearly with bandwidth, and as the square of the number of stations. The exploration of scalable approaches to address future EHT correlation needs is underway (Gill et al. 2019). The combination of these efforts is aimed at the substantial expansion of aperture plane sampling of the EHT and improved imaging.

To significantly improve angular resolution requires development in different directions. Planned extension of the EHT to operation at 0.87 mm wavelength, a standard receiving band at many facilities, would increase the angular resolution of the array by $\sim$40%. VLBI tests at 0.87 mm are underway and this capability is expected at a subset of EHT sites over the next 3–5 yr (see Table 4). An alternate approach to increased angular resolution is to deploy EHT antennas in space where baseline length is not limited to the diameter of the Earth, potentially allowing horizon-scale imaging of additional SMBH candidates (Johannsen et al. 2012). Space platforms also offer the possibility of rapidly sampling the Fourier plane, thereby opening the potential for dynamical imaging of black hole accretion and outflow processes (D. Palumbo et al. 2019, in preparation; F. Roelofs et al. 2019, in preparation; M. Shea et al. 2019, in preparation).

EHT observation strategies continue to be refined. With further development of remote monitoring and control tools, the EHT can explore triggering observations during the best conditions throughout the year outside the current March–April window. One consequence of this flexibility would be that observations for Sgr A* and M87 could be optimized separately in a given observing cycle, instead of grouped together as they are currently. Distributing the observations over a larger portion of the year also increases the likelihood that the EHT would detect emission transients or flaring on intermediate timescales should they occur, and affords, in general, the opportunity to study M87 on timescales that correspond to the expected ISCO period of order one to several weeks (Table 1). Not all EHT sites may be able to participate in such flexible campaigns, but even a subset of the array, especially if augmented with many smaller dishes, could provide useful observations for variability studies.

## 7. Conclusion

The goal of the EHT project is to observe SMBHs with spatial and temporal resolution that permits imaging on the scale of the lensed photon orbit, and the study of dynamics on commensurate light-crossing timescales. Leading up to the detection of event-horizon-scale structure in Sgr A* in 2008, and now a decade afterward, the development of EHT instrumentation proceeded through a series of systems that supported increasingly ambitious observations. These observing campaigns led to precursor scientific results that motivated a strategy of building an imaging array: one with baselines long enough to resolve horizon-scale structure, that included enough geographical sites to sufficiently sample the Fourier plane, and had the sensitivity required to detect compact, low flux density targets. Technology maturation, fueled by industry-driven trends, was focused primarily on improving the bandwidth of observations by increasing digitization and recording rates. Over a period of 10 yr, EHT data capture rates increased from 4 to 64 Gbps, allowing for observations that take full advantage of the capabilities of modern millimeter receivers. The parallel development of phased array technologies and investment in site-specific infrastructure (see the Appendix) have now led to a global EHT array with the capability to address its science goals.

Had the EHT project needed to build a VLBI array from the ground up, the cost would have been prohibitive. By developing new systems and infrastructure that allowed for the use of existing telescopes and facilities, most of which were not originally conceived to operate as VLBI elements, the EHT has created a purpose-built array with unique capability at modest cost.

The authors of this Letter thank the following organizations and programs: the Academy of Finland (projects 274477, 284495, 312496); the Advanced European Network of E-infrastructures for Astronomy with the SKA (AENEAS) project, supported by the European Commission Framework Programme Horizon 2020 Research and Innovation action under grant agreement 731016; the Alexander von Humboldt Stiftung; the Black Hole Initiative at Harvard University, through a grant (60477) from the John Templeton Foundation; the China Scholarship Council; Comisión Nacional de Investigación Científica y Tecnológica (CONICYT, Chile, via PIA ACT172033, Fondecyt 1171506, BASAL AFB-170002, ALMA-conicyt 31140007); Consejo Nacional de Ciencia y Tecnología (CONACYT, Mexico, projects 104497, 275201, 279006, 281692); the Delaney Family via the Delaney Family John A. Wheeler Chair at Perimeter Institute; Dirección General de Asuntos del Personal Académico–Universidad Nacional Autónoma de México (DGAPA–UNAM, project IN112417); the European Research Council Synergy Grant "BlackHoleCam: Imaging the Event Horizon of Black Holes" (grant 610058); the Generalitat Valenciana postdoctoral grant APOSTD/2018/177; the Gordon and Betty Moore Foundation (grants GBMF-3561, GBMF-5278); the Istituto Nazionale di Fisica Nucleare (INFN) sezione di Napoli, iniziative specifiche TEONGRAV; the International Max Planck Research School for Astronomy and Astrophysics at the Universities of Bonn and Cologne; the Jansky Fellowship program of the National Radio Astronomy Observatory (NRAO); the Japanese Government (Monbukagakusho: MEXT) Scholarship; the Japan Society for the Promotion of Science (JSPS) Grant-in-Aid for JSPS Research Fellowship (JP17J08829); JSPS Overseas Research Fellowships; the Key Research Program of Frontier Sciences, Chinese Academy of Sciences (CAS, grants QYZDJ-SSW-SLH057, QYZDJ-SSW-SYS008); the Leverhulme Trust Early Career Research Fellowship; the Max-Planck-Gesellschaft (MPG); the Max Planck Partner Group of the MPG and the CAS; the MEXT/JSPS KAKENHI (grants

18KK0090, JP18K13594, JP18K03656, JP18H03721, 18K03709, 18H01245, 25120007); the MIT International Science and Technology Initiatives (MISTI) Funds; the Ministry of Science and Technology (MOST) of Taiwan (105-2112-M-001-025-MY3, 106-2112-M-001-011, 106-2119-M-001-027, 107-2119-M-001-017, 107-2119-M-001-020, and 107-2119-M-110-005); the National Aeronautics and Space Administration (NASA, Fermi Guest Investigator grant 80NSSC17K0649); the National Institute of Natural Sciences (NINS) of Japan; the National Key Research and Development Program of China (grant 2016YFA0400704, 2016YFA0400702); the National Science Foundation (NSF, grants AST-0096454, AST-0352953, AST-0521233, AST-0705062, AST-0905844, AST-0922984, AST-1126433, AST-1140030, DGE-1144085, AST-1207704, AST-1207730, AST-1207752, MRI-1228509, OPP-1248097, AST-1310896, AST-1312651, AST-1337663, AST-1440254, AST-1555365, AST-1715061, AST-1615796, AST-1614868, AST-1716327, OISE-1743747, AST-1816420); the Natural Science Foundation of China (grants 11573051, 11633006, 11650110427, 10625314, 11721303, 11725312, 11873028, 11873073, U1531245, 11473010); the Natural Sciences and Engineering Research Council of Canada (NSERC, including a Discovery Grant and the NSERC Alexander Graham Bell Canada Graduate Scholarships-Doctoral Program); the National Youth Thousand Talents Program of China; the National Research Foundation of Korea (grant 2015-R1D1A1A01056807, the Global PhD Fellowship Grant: NRF-2015H1A2A1033752, and the Korea Research Fellowship Program: NRF-2015H1D3A1066561); the Netherlands Organization for Scientific Research (NWO) VICI award (grant 639.043.513) and Spinoza Prize SPI 78-409; the New Scientific Frontiers with Precision Radio Interferometry Fellowship awarded by the South African Radio Astronomy Observatory (SARAO), which is a facility of the National Research Foundation (NRF), an agency of the Department of Science and Technology (DST) of South Africa; the Onsala Space Observatory (OSO) national infrastructure, for the provisioning of its facilities/observational support (OSO receives funding through the Swedish Research Council under grant 2017-00648) the Perimeter Institute for Theoretical Physics (research at Perimeter Institute is supported by the Government of Canada through the Department of Innovation, Science and Economic Development Canada and by the Province of Ontario through the Ministry of Economic Development, Job Creation and Trade); the Russian Science Foundation (grant 17-12-01029); the Spanish Ministerio de Economía y Competitividad (grants AYA2015-63939-C2-1-P, AYA2016-80889-P); the State Agency for Research of the Spanish MCIU through the "Center of Excellence Severo Ochoa" award for the Instituto de Astrofísica de Andalucía (SEV-2017-0709); the Toray Science Foundation; the US Department of Energy (USDOE) through the Los Alamos National Laboratory (operated by Triad National Security, LLC, for the National Nuclear Security Administration of the USDOE (Contract 89233218CNA000001)); the Italian Ministero dell'Istruzione Università e Ricerca through the grant Progetti Premiali 2012-iALMA (CUP C52I13000140001); the European Union's Horizon 2020 research and innovation programme under grant agreement No 730562 RadioNet; ALMA North America Development Fund Chandra TM6-17006X.

This work used the Extreme Science and Engineering Discovery Environment (XSEDE), supported by NSF grant ACI-1548562, and CyVerse, supported by NSF grants DBI-0735191, DBI-1265383, and DBI-1743442. XSEDE Stampede2 resource at TACC was allocated through TG-AST170024 and TG-AST080026N. XSEDE JetStream resource at PTI and TACC was allocated through AST170028. The simulations were performed in part on the SuperMUC cluster at the LRZ in Garching, on the LOEWE cluster in CSC in Frankfurt, and on the HazelHen cluster at the HLRS in Stuttgart. This research was enabled in part by support provided by Compute Ontario (http://computeontario.ca), Calcul Quebec (http://www.calculquebec.ca) and Compute Canada (http://www.computecanada.ca).

The EHT Collaboration is indebted to the following people for their contributions to the success of the 2017 EHT observations: Joost Adema, Claudio Agurto, Hector Alarcon, Jonathan Antognini, Juan Pablo Araneda, Oriel Arriagada, Jorge Avarias, Amit Bansod, Denis Barkats, Emilio Barrios, Alain Baudry, Alessandra Bertarini, Andy Biggs, Alan Bridger, Michel Caillat, Michael Cantzler, Patricio Caro, John Carpenter, Jorge Castillo, Miroslaw Ciechanowicz, Stuartt Corder, Antonio Cordoba, Pierre Cox, Faviola Cruzat, Mauricio Cárdenas, Itziar De Gregorio, Bill Dent, Carlos Duran, Ray Escoffier, Soledad Fuica, Enrique Garcia, Juan Carlos Gatica, Juan Pablo Gil, Brian Glendenning, Edouard Gonzales, Joe Greenberg, Rolf Güsten, Rob de Haan-Stijkel, Hayor Haas, Hayo Hase, Christian Herrera, Daniel Herrera, Richard Hills, Rafael Hiriart, Arturo Hoffstadt, Jorge Ibsen, Christophe Jacques, Jan Johansson, Chris Kendall, Jeff Kern, Thomas Klein, Rudiger Kneissel, Albert Koops, James Lamb, Bernhard Lopez, Cristian Lopez, Robert Lucas, Felipe MacAuliffe, Gianni Marconi, Lorenzo Martinez-Conde, Mary Mayo, Mark McKinnon, Francisco Montenegro-Montes, Rolf Märtens, Lars-Åke Nyman, Rodrigo Olivares, Hans Olofsson, Victor Pankratius, Miroslav Pantaleev, Manuel Parra, Rodrigo Parra, J Perez, Dave Pernic, Juan Pablo Pérez-Beaupuits, Jorge Ramírez, William Randolph, Anthony Remijan, Johnny Reveco, Rachel Rosen, Norman Saez, Ana Salinas, Jorge Santana, Jorge Sepulveda, Tzu-Chiang Shen, Bill Shillue, Ruben Soto, Tomas Staig, Tsuyushi Swada, Satoko Takahashi, Claudia Tapia, Karl Torstensson, Robert Treacy, Gino Tuccari, Paulina Venegas, Eric Villard, Joseph Weber, Nick Whyborn, Gundolf Wieching, Michael Wunderlich.

We thank the staff at the participating observatories, correlation centers, and institutions for their enthusiastic support.

This Letter makes use of the following ALMA data: ADS/JAO.ALMA#2016.1.01154.V. ALMA is a partnership of the European Southern Observatory (ESO; Europe, representing its member states), NSF, and National Institutes of Natural Sciences of Japan, together with National Research Council (Canada), Ministry of Science and Technology (MOST; Taiwan), Academia Sinica Institute of Astronomy and Astrophysics (ASIAA; Taiwan), and Korea Astronomy and Space Science Institute (KASI; Republic of Korea), in cooperation with the Republic of Chile. The Joint ALMA Observatory is operated by ESO, Associated Universities, Inc. (AUI)/NRAO, and the National Astronomical Observatory of Japan (NAOJ). The NRAO is a facility of the NSF operated under cooperative agreement by AUI. APEX is a collaboration between the Max-Planck-Institut für Radioastronomie (Germany), ESO, and the Onsala Space Observatory (Sweden). The GLT is supported by

the Academia Sinica (Taiwan), the MOST (Taiwan) grants (99-2119-M-001-002-MY4, 103-2119-M-001-010-MY2, 106-2119-M-001-013, 106-2923-M-001-005, and 107-2923-M-001-002), and the Smithsonian Institution. The GLT project thanks the National Chun-Shan Institute of Science and Technology (Taiwan) for their strong support, NAOJ for their support in receiver instrumentation, the NSF Office of Polar Programs for effective support in logistics, and the United States Air Force, 821st Air Base Group, Thule Air Base, Greenland for use of the site and access to the base and logistics chain. The GLT project appreciates the development work done by Vertex Antennentechnik GmbH and ADS international, and also thanks Atunas, an outdoor wear company in Taiwan, for the arctic clothing. NRAO and the Massachusetts Institute of Technology Haystack Observatory supported the acquisition of the ALMA–North America prototype antenna and its repurposing for deployment in Greenland. The NOEMA observatory at the Plateau de Bure, France, and the IRAM 30-m telescope on Pico Veleta, Spain, are operated by IRAM and supported by CNRS (Centre National de la Recherche Scientifique, France), MPG (Max-Planck-Gesellschaft, Germany) and IGN (Instituto Geográfico, Nacional, Spain). The SMA is a joint project between the SAO and ASIAA and is funded by the Smithsonian Institution and the Academia Sinica. The JCMT is operated by the East Asian Observatory on behalf of the NAOJ, ASIAA, and KASI, as well as the Ministry of Finance of China, Chinese Academy of Sciences, and the National Key R&D Program (No. 2017YFA0402700) of China. Additional funding support for the JCMT is provided by the Science and Technologies Facility Council (UK) and participating universities in the UK and Canada. The LMT project is a joint effort of the Instituto Nacional de Astrófisica, Óptica, y Electrónica (Mexico) and the University of Massachusetts at Amherst (USA).

The SMT and KP are operated by the Arizona Radio Observatory, a part of the Steward Observatory of the University of Arizona, with financial support of operations from the State of Arizona and financial support for instrumentation development from the NSF. Partial SPT support is provided by the NSF Physics Frontier Center award (PHY-0114422) to the Kavli Institute of Cosmological Physics at the University of Chicago (USA), the Kavli Foundation, and the GBMF (GBMF-947). The SPT hydrogen maser was provided on loan from the GLT, courtesy of ASIAA. The SPT is supported by the National Science Foundation through grant PLR-1248097. Partial support is also provided by the NSF Physics Frontier Center grant PHY-1125897 to the Kavli Institute of Cosmological Physics at the University of Chicago, the Kavli Foundation and the Gordon and Betty Moore Foundation grant GBMF 947.

The EHTC has received generous donations of FPGA chips from Xilinx Inc., under the Xilinx University Program. The EHTC has benefited from technology shared under open-source license by the Collaboration for Astronomy Signal Processing and Electronics Research (CASPER). The EHT project is grateful to T4Science and Microsemi for their assistance with Hydrogen Masers. This research has made use of NASA's Astrophysics Data System. We gratefully acknowledge the support provided by the extended staff of the ALMA, both from the inception of the ALMA Phasing Project through the observational campaigns of 2017 and 2018. We would like to thank A. Deller and W. Brisken for EHT-specific support with the use of DiFX. We acknowledge the significance that

# Appendix
## Specific Details of Participating Facilities in EHT Observations

One of the technical challanges of the EHT array is that it is assembled from existing telescope facilities that differ to a greater or lesser extent in their technical characteristics and readiness for VLBI operations. As a result, EHT instrumentation installed at each site requires customization. This appendix provides information on the specifics of each site in turn, not general to the array overall.

### A.1. ALMA (5100 m Altitude)

The ALMA, located on the Chajnantor plain in Chile, is an international partnership between Europe, the United States, Canada, Japan, South Korea, Taiwan, and Chile (Wootten & Thompson 2009). The connected-element interferometer consists of fifty 12 m and twelve 7 m antennas, and is supplemented with four 12 m total-power antennas. Reflector surface accuracy is better than Mark 6 VLBI recorders was installed. $25\,\mu m$ rms.

The ALMA Phasing Project (APP) was an international effort to produce an antenna phasing system leveraging the superb (sub)millimeter capabilities and large collecting area of ALMA for VLBI applications (Doeleman 2010; Matthews et al. 2018). The APP added to the hardware and software already available at the ALMA site for connected-element interferometry the additional components necessary for coherently summing the ALMA antennas and recording VLBI data products. The rubidium clock that was formerly used to generate the reference 5 MHz LO tone was replaced with a hydrogen maser. Phasing interface cards were added to the ALMA Baseline Correlator to serve as the VLBI backend. A fiber link system was built and deployed to transport the phased-sum signal from the ALMA high-elevation site to the ALMA Operations Support Facility at an elevation of about 2900 m, where a set of Mark 6 VLBI recorders was installed. Numerous software enhancements were also required, including the implementation of an ALMA VLBI Observing Mode (VOM) and a phase solver to calculate the phases needed to adjust each of the ALMA antennas to allow coherent summation of their signals. The ALMA phasing system treats the phased sum as another input to the Baseline Correlator, so the phased-sum signal can be correlated with individual antenna signals. Standard ALMA single-field interferometric data are generated as a matter of course during VOM operation, and ALMA-only data products are archived as usual. When all antennas are included in the phased sum, ALMA is equivalent to an effective ∼88 m diameter aperture for VLBI operations. Details are described in Matthews et al. (2018) and Goddi et al. (2019).

### A.2. APEX (5100 m Altitude)

APEX is a 12 m antenna located on the Chajnantor plain in Chile about 2 km from the ALMA site (Güsten et al. 2006). It was built by VERTEX Antennentechnik from the ALMA prototype development and inaugurated in 2005. It was outfitted and operated by the Max-Planck-Institut für Radio-astronomie in Bonn (50%), the Onsala Space Observatory (23%), and the European Southern Observatory (27%) as the first telescope on the Chajnantor plain. The first VLBI fringes were obtained with APEX in 2012 at 230 GHz (Wagner et al. 2015).

The reflector surface accuracy is 17 $\mu$m rms, which gives good efficiency at frequencies exceeding 1 THz, and the telescope is correspondingly equipped with receivers spanning 200 GHz–1.4 THz. The surface panels and subreflector were replaced in early 2018 to improve the surface to 10 $\mu$m rms and so further raise the efficiency at the highest frequencies. At 230 GHz the elevation gain curve is flat to about 3% due to the high structural rigidity of the antenna. Pointing accuracy is better than 2 arcsec rms, and tracking accuracy is typically below 1 arcsec rms depending on wind conditions. The beam FWHM is 27.1 arcsec at 228.1 GHz.

The maser 10 MHz reference is transported to the front-end synthesizers and backend equipment on four low temperature-coefficient, double-shielded 85 m length cables. The cables are wrapped in thermal insulation and firmly fixed to support structures in the receiver cabin to minimize phase instabilities resulting from temperature changes or telescope movement. No round-trip phase stabilization is used. The front-end synthesizers (first LO and tone generation) each have their own frequency reference cable from the maser. References for the backend equipment are supplied through a distributor because the frequencies generated are much lower than for the front-end. The R2DBE and another VLBI backend called the DBBC are located in the antenna close to the receiver (15 m cable length) for low loss transmission of the dual-polarization dual-sideband 4–12 GHz receiver IF output. The Mark 6 recorders are located in a control container about 50 m from the telescope. The backends transmit eight 8 Gbps VDIF-format data streams on fiber with short-range 10 GbE 850 nm SFP+ fiber transceivers to the recorders.

### A.3. GLT (100 m Altitude Currently, 3200 m Altitude Planned)

The GLT (Inoue et al. 2014) was originally the 12 m ALMA-North America prototype antenna (Mangum et al. 2006) located at the Very Large Array (VLA) site near Socorro, New Mexico. The antenna was awarded to the Smithsonian Astrophysical Observatory (SAO) in 2011, representing the partnership of the SAO and Academia Sinica Institute of Astronomy (ASIAA). The antenna was retrofitted and rebuilt for operation in the extreme arctic conditions of northern Greenland. Presently located at a temporary site on Thule Air Base, Greenland, the telescope is equipped with a new set of submillimeter receivers operating at 86, 230, and 345 GHz. In 2018, the GLT dish surface underwent iterative adjustments to improve surface accuracy, eventually achieving better than 40 $\mu$m rms. The pointing accuracy is about 3–5 arcsec during 230 GHz observations. The telescope is planned to be relocated to Greenland Summit, where the atmosphere provides excellent transparency for submm and THz observations (Matsushita et al. 2017).

The GLT has a complete VLBI backend. In 2018 January, fringes were detected between the GLT and ALMA, and in 2018 April, the GLT participated in EHT observations (Chen et al. 2018). The maser and associated equipment are located in a dedicated trailer approximately 40 m from the telescope. The maser 10 MHz is fed through a phase-stable coaxial cable to an adjacent synthesizer that generates the first LO (18–31 GHz) reference for the receivers. This first LO reference along with a separate maser reference of 100 MHz are combined and transmitted to the antenna through a single-mode fiber and used to phase-lock all the associated instruments in the receiver cabin (Kubo et al. 2018). The BDCs and R2DBEs are located in the receiver cabin and are interconnected using short segments of phase-stable coaxial cables. The Mark 6 recorders are located in the VLBI trailer approximately 35 m from the telescope. Communications between the R2DBEs and the Mark 6 recorders are via a 10 GbE optical network. The current system supports a throughput of 64 Gbps to the recorders.

### A.4. JCMT (4100 m Altitude)

The JCMT is located on Maunakea in Hawai'i. It was dedicated in 1987 and was operated by a consortium (the United Kingdom, Canada, and the Netherlands) until 2015. Since then, the East Asia Observatory has operated the JCMT with funding from Taiwan, China, Japan, Korea, the United Kingdom, and Canada. The reflector is a 15 m Cassegrain design and has a surface accuracy of about 25 $\mu$m rms. The telescope is enclosed in a co-rotating dome and is protected by a 0.3 mm thick GoreTex$^{TM}$ windblind that is not removed for observing. The atmospheric opacity is measured with a line-of-sight 183 GHz radiometer.

The telescope participated in the VLBI observations that in 2007 detected the horizon-scale emission in Sgr A* (Doeleman et al. 2008) and has successfully participated in all EHT observing since then. This was done in cooperation with the SMA, which generated the phase-stable frequency reference signals and recorded the IF, utilizing the extended SMA (eSMA) infrastructure that enables the JCMT to join the SMA array as a connected element. From the 2019 VLBI campaign the JCMT will use a locally installed VLBI backend for recording, and will be equipped with a new receiver outfitted with ALMA 230 and 345 GHz mixers at 1.3 and 0.87 mm built by ASIAA.

### A.5. Institut de Radioastronomie Millimétrique: 30 m (PV, 2900 m Altitude)

The IRAM 30 m telescope is located on Pico Veleta in the Spanish Sierra Nevada near Granada. The telescope was built in the early 1980s (Baars et al. 1987). The first VLBI fringes at $\sim$1.3 mm were obtained in 1995 (Greve et al. 1995) with one antenna of the Plateau de Bure interferometer, and since 1996 the PV 30 m has participated in the regular 3 mm VLBI CMVA/GMVA biannual runs.

The telescope surface has been adjusted to an accuracy of about 55 $\mu$m rms using phase-coherent holography with geostationary satellites. This yields a 230 GHz aperture efficiency of 47% at the optimal elevation of 50°. The gain elevation curve drops to about 35 % at elevations of 20° and 80°. The blind pointing accuracy is about 3 arcsec. The tracking accuracy is better than 1 arcsec for wind speeds below about 10 m s$^{-1}$. Active thermal control of the telescope backstructure allows for efficient operation around the clock, and a de-icing system enables quick startup after winter storms. A 225 GHz radiometer continuously monitors the sky opacity at a fixed azimuth. Since 2018 October, the 30 m telescope is connected via a 1 Gbps fiber link to IRAM/Granada and the Spanish research network RedIRIS.

The PV 30 m is currently equipped with two heterodyne receivers. One of these, the Eight MIxer Receiver (EMIR)

installed in 2009, has eight sideband-separating (2SB) mixers. The EMIR has 8 GHz of bandwidth per sideband covering the atmospheric windows in 3, 2, 1.3, and 0.87 mm wavelengths in dual-polarization. The EHT uses the EMIR 1.3 and 0.87 mm bands. Conversion from linear to circular polarization is done by quarter-wave plates, which are installed in a carousel to quickly switch between non-VLBI and VLBI observations. The IF signal from the receiver cabin is transported to a temperature-controlled backend room over 100 m coaxial cables. A new hydrogen maser was installed in 2018 January and then used for the GMVA and EHT sessions in 2018 April. For the EHT, four R2DBEs deliver 64 Gbps (two polarizations, two sidebands) to four Mark 6 recorders (see Figure 8).

### A.6. IRAM: NOEMA (2600 m Altitude)

NOEMA is located in the French Alps and is the largest millimeter-wave facility in the northern hemisphere. With twelve 15 meter antennas, 2SB receivers with 8 GHz IF per sideband and polarization, and baselines out to 1.7 km, NOEMA will operate at 10 times the sensitivity and 4 times the spatial resolution of its predecessor, the former Plateau de Bure Interferometer (Guilloteau et al. 1992).

The NOEMA Phase 1 extension program was completed in 2018 September with the delivery of four additional antennas to the existing six, each with a surface accuracy of 35 $\mu$m rms. The dual-polarization 2SB receivers operate in the 3, 2, and 1.3 mm wavelength bands, with noise temperatures of, respectively, 35 K, 40 K, and 50 K in each band. Signal processing is done using an FPGA-based FFX correlator with flexible configurations modes. Beginning in 2019, NOEMA will have a phased-array mode with an initial processing bandwidth of 2 × 2 GHz (16 Gbps), eventually growing to 128 Gbps. The NOEMA Phase 2 extension program (2019) includes the construction of two additional antennas, the delivery of a second correlator for dual-band operation (2 × 32 GHz), a receiver upgrade to perform full-array observations up to 373 GHz (0.8 mm), and a baseline extension to 1.7 km for high-fidelity imaging down to a spatial resolution of 0.1 arcsec.

### A.7. Kitt Peak 12 m (KP, 1900 m Altitude)

The telescope on Kitt Peak was acquired by the University of Arizona in 2013 March. The 12 m dish was a prototype antenna (Mangum et al. 2006) made for the ALMA project by the Alcatel/European Industrial Engineering consortium. It was relocated from the VLA site in New Mexico to the enclosure of the previous NRAO 12 m telescope on Kitt Peak, Arizona, in 2013 November. The telescope is operated by the Arizona Radio Observatory (ARO), a division of the Steward Observatory of the University of Arizona. First light was observed in 2014 October. The surface was adjusted to an rms figure error of 16 $\mu$m in 2018 September using photogrammetry. A 225 GHz radiometer measures atmospheric opacity.

VLBI observations will make use of a newly developed multi-band receiver (1.3–4 mm), which incorporates ALMA Band 3 and Band 6 mixers for the 3 and 1.3 mm bands. All bands sample both polarizations, and the 1.3 mm band will be converted to circular polarizations through a quartz quarter-wave plate. The receiver is under commissioning and the telescope is scheduled to participate in the next scheduled EHT observations. The master reference for this site is located within the domed enclosure adjacent to the computer room that houses the VLBI backend electronics.

### A.8. LMT (4600 m Altitude)

The LMT is situated at the summit of Volcán Sierra Negra in Central Mexico. It is jointly operated by the Instituto Nacional de Astrofísica, Óptica y Electrónica (INAOE) and the University of Massachusetts (UMass). In 2017, two additional rings of precision surface panels were installed to enlarge the LMT primary mirror from 32.5 to 50 m. The primary mirror is an active surface comprised of 180 segments steered by 720 actuators to correct thermal and mechanical distortions. The large collecting area and a central geographical location with respect to other EHT sites make the LMT particularly important for array sensitivity and imaging fidelity.

Initial technical work to develop 3 mm wavelength VLBI at the LMT was the product of a multi-year collaboration between UMass, INAOE, SAO, the Massachusetts Institute of Technology Haystack Observatory, the Universidad Nacional Autónoma de México (UNAM), and NRAO. The first 3 mm observations were carried out in 2013 April with the facility Redshift Search Receiver (Erickson et al. 2007), a broadband dual-polarization, dual-beam receiver that acted as the front end. Those observations were performed with a VLBI backend using a high-precision quartz oscillator frequency reference running at 2 Gbps recording rate. Subsequent integration of a hydrogen maser and further testing in 2014 led to first VLBI science observations at 3 mm wavelength in 2015 (Ortiz-León et al. 2016).

In 2014, an interim special-purpose VLBI 1.3 mm wavelength receiver was assembled using mixers from the CARMA observatory and resourced through a collaboration including the institutes above and the University of California, Berkeley. This receiver was a single-pixel, dual-polarization, double-sideband receiver. It was deployed in 2015 for commissioning along with a 32 Gbps VLBI system (Figure 4), both of which were used for the 2017 April EHT observations. In early 2018, a new dual-polarization, sideband-separating 230 GHz receiver, using ALMA mixers, was delivered by UMass and installed along with an upgrade of the VLBI system to 64 Gbps capability.

The hydrogen maser at the LMT is located in a temperature-stable environment deep in the apex cone. The maser reference is brought to a 10 MHz distribution system in the VLBI backend room. The backend room is located one level below the receiver cabin and houses the 2.048 GHz synthesizer, the 1 PPS distribution system, BDCs, R2DBEs, and the Mark 6 recorders.

### A.9. SMA (4100 m Altitude)

The SMA is an eight-element radio interferometer located atop Maunakea operated by SAO and ASIAA (Ho et al. 2004). The 6 m dishes are configurable with baselines as long as 509 m, which produces a synthesized beam of about 1 arcsec width at 230 GHz, and sub-arcsec width at 345 GHz. The SMA has participated in EHT observations since 2006. In 2006 and 2007, the SMA provided a maser reference tuning signal to the Caltech Submillimeter Observatory and the JCMT, respectively. In 2009, the SMA contributed collecting area to the EHT through the Phased Array Recording Instrument for Galactic Event Horizon Studies (PhRInGES), a 4 Gbps phased-array system.

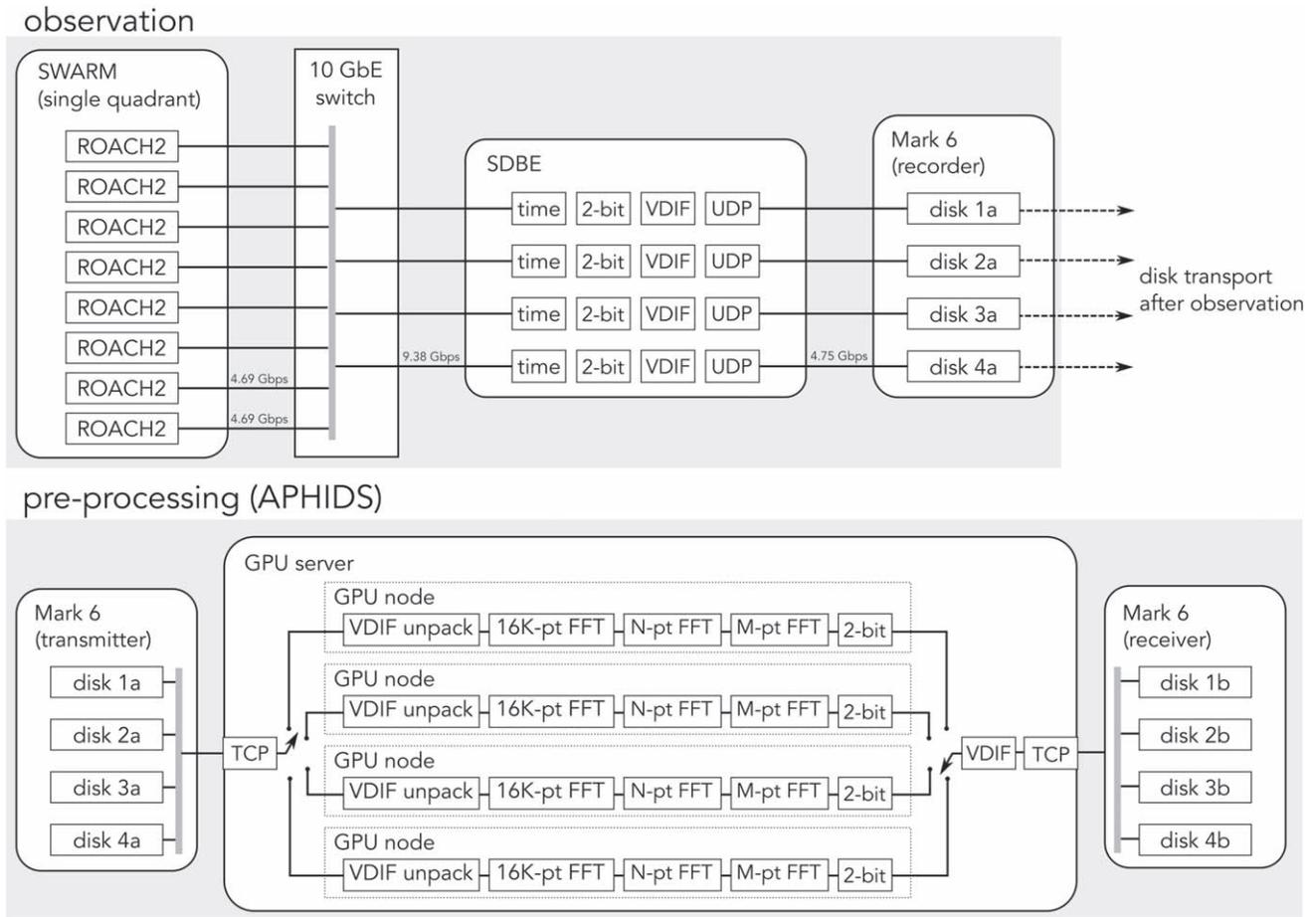

**Figure 14.** Block diagram demonstrating the VLBI data pipeline, from SWARM to correlatable data in VDIF format. The SDBE is integrated with SWARM and does the real-time processing necessary to interface with the on-site Mark 6 during an observation. After observing, the data is preprocessed offline in APHIDS prior to correlation with data from other sites.

During EHT observations, the SMA uses its own correlator, known as the SMA Wideband Astronomical ROACH2 Machine (SWARM), instead of the R2DBE. SWARM is a combined correlator and phased array that was initially deployed in 2014 and built out to its full 32 GHz bandwidth by 2017 (Primiani et al. 2016). SWARM supports VLBI through a built-in beamformer based on a signal-noise decomposition of the correlator output while observing the target source (Young et al. 2016). Phasing the array requires tracking all sources of delay, including fluctuations in water vapor concentration in the atmosphere. The SWARM phasing system is equipped with a real-time phasing solver that continually updates the beamforming weights over the course of the observation to compensate for these variable delays, which manifest as variable phase errors in each antenna. Because the phased array capability is used to observe sources that are unresolved on baselines within the array, the corrective beamformer weights can be computed by extracting from the correlator output that contribution associated with a point-like source. Furthermore, as the weights are applied to the signal from each antenna before computing cross-correlations between antenna pairs, the solution obtained from the correlator output for a particular integration period can also be used to calculate the average phasing efficiency over that same period. In addition to solving and applying corrective phasing in real time, the system produces an estimate of the phasing efficiency, $\eta_\phi$,

$$\eta_\phi = \left| \sum_i w_i \right|^2 \Big/ \left( \sum_i |w_i| \right)^2, \qquad (2)$$

where $w_i$ is the complex-valued weight applied to antenna i. See Young et al. (2016)[144] for a more detailed discussion of the phased array and performance assessment thereof.

While the SWARM correlator architecture lends itself conveniently to phased-array processing, it also presents several challenges for integrating the instrument into a VLBI array. First, the beamformer processing of 2 GHz bands is spread across eight ROACH2 devices. These parallel data streams must be collected and formatted in real time in order to record to the Mark 6, which is done using a custom FPGA device called the SWARM Digital Backend (SDBE). The data are received on four of the eight 10 GbE ports on the SDBE. The packets are time-stamped, the frequency domain samples are quantized from 4-bit complex to 2-bit complex, the packets are formatted with VDIF headers, and then transported over Ethernet to the Mark 6. As the beamformer data packets are relatively small, several of these packets are bundled into each UDP packet to reduce the interrupt rate on the Mark 6 and prevent packet loss. The second complication to using SWARM in VLBI is that the data are in complex frequency domain representation as opposed to the time-domain format used by the majority of stations in the EHT, and the data are sampled at a different rate than the single-dish EHT stations: 4.576 Gsps compared to 4.096 Gsps at single-dish sites. Prior to the correlation of SWARM data with data from other sites, an offline inversion and resampling program known as the Adaptive Phased-array and Heterogeneous Interpolating Downsampler for SWARM (APHIDS) is used to preprocess the SMA data. Development of a real-time preprocessing system to be integrated into the SDBE is currently in progress. The SMA VLBI pipeline is illustrated in Figure 14.

------

[144] See also the related SMA Memo 163 at https://www.cfa.harvard.edu/sma/memos/163.pdf.

### A.10. SMT (3200 m Altitude)

The SMT, located on Mount Graham, Arizona, was built by the Max-Planck-Institut für Radioastronomie in 1993, and is now operated and maintained by the ARO (Baars et al. 1999). Work to incorporate the telescope in mm-wavelength VLBI experiments began in the late 1990s (Doeleman et al. 2002; Krichbaum et al. 2002). In 2007, the SMT participated in the observations that detected horizon-scale emission in Sgr A* (Doeleman et al. 2008). The 10 m telescope is mounted in a co-rotating enclosure that minimizes the effects of insolation and wind during observations. The telescope surface accuracy is ∼15 μm rms and the telescope pointing is typically accurate to better than 1.5 arcsec rms. Due to weak diurnal variations in water vapor and the protection of the telescope enclosure, the SMT can observe 24 hr per day. Atmospheric opacity is monitored at this site with a tipping 225 GHz radiometer that is mounted to the rotating dome.

The SMT is equipped with dual-polarization heterodyne receivers covering the atmospheric windows from ∼200 to 700 GHz. For VLBI observations, the 1.3 mm receiver system uses prototype ALMA sideband-separating mixers, with polarization splitting achieved via a room-temperature wire grid, and circular polarization induced through a quartz quarter-wave plate. The SMT also has a double-sideband receiver for observations at 0.87 mm; a replacement sideband-separating receiver for this band is in development at the University of Arizona. The 64 Gbps EHT backend hardware is mounted in the receiver cabin. The master reference for the SMT is a hydrogen maser mounted at the base of the telescope pier in a temperature-stabilized closet.

### A.11. SPT (2800 m Altitude)

The SPT is a 10 m diameter mm-wavelength telescope located at the geographic South Pole (Carlstrom et al. 2011) and is operated by a collaboration led by the University of Chicago. The off-axis Gregorian design is optimized for observations of the cosmic microwave background with wide-field, multi-color bolometer cameras. The primary reflector surface accuracy is 20 μm rms.

EHT observations with the SPT are managed by the University of Arizona. The standard observing campaigns occur during the austral fall when the South Pole is inaccessible. VLBI setup and testing is performed by a two-person SPT winter-over team, with remote supervision from Arizona. EHT observing at the SPT employs a dual-frequency 230/345 GHz VLBI receiver (Kim et al. 2018a), which was constructed at the University of Arizona. The receiver was first installed in late 2014 and first operated in 2015 January (Kim et al. 2018b). The receiver was reinstalled in an upgraded receiver cabin in 2016 in preparation for the 2017 EHT campaign. To illuminate the VLBI receiver, an alternate secondary mirror is installed ahead of prime focus to form a Cassegrain telescope, which relays the light to an ellipsoidal tertiary mirror atop the receiver. The full VLBI signal chain includes a hydrogen maser and coherence monitoring loops to account for the unusual range of environmental temperatures at the site. Mark 6 data modules are shipped after flights resume in the austral spring season, around November 1.

### ORCID iDs

Kazunori Akiyama  https://orcid.org/0000-0002-9475-4254
Antxon Alberdi  https://orcid.org/0000-0002-9371-1033
Rebecca Azulay  https://orcid.org/0000-0002-2200-5393
Anne-Kathrin Baczko  https://orcid.org/0000-0003-3090-3975
Mislav Baloković  https://orcid.org/0000-0003-0476-6647
John Barrett  https://orcid.org/0000-0002-9290-0764
Lindy Blackburn  https://orcid.org/0000-0002-9030-642X
Katherine L. Bouman  https://orcid.org/0000-0003-0077-4367
Geoffrey C. Bower  https://orcid.org/0000-0003-4056-9982
Christiaan D. Brinkerink  https://orcid.org/0000-0002-2322-0749
Roger Brissenden  https://orcid.org/0000-0002-2556-0894
Silke Britzen  https://orcid.org/0000-0001-9240-6734
Avery E. Broderick  https://orcid.org/0000-0002-3351-760X
Do-Young Byun  https://orcid.org/0000-0003-1157-4109
Andrew Chael  https://orcid.org/0000-0003-2966-6220
Chi-kwan Chan  https://orcid.org/0000-0001-6337-6126
Shami Chatterjee  https://orcid.org/0000-0002-2878-1502
Ilje Cho  https://orcid.org/0000-0001-6083-7521
Pierre Christian  https://orcid.org/0000-0001-6820-9941
John E. Conway  https://orcid.org/0000-0003-2448-9181
Geoffrey B. Crew  https://orcid.org/0000-0002-2079-3189
Yuzhu Cui  https://orcid.org/0000-0001-6311-4345
Jordy Davelaar  https://orcid.org/0000-0002-2685-2434
Mariafelicia De Laurentis  https://orcid.org/0000-0002-9945-682X
Roger Deane  https://orcid.org/0000-0003-1027-5043
Jessica Dempsey  https://orcid.org/0000-0003-1269-9667
Gregory Desvignes  https://orcid.org/0000-0003-3922-4055
Jason Dexter  https://orcid.org/0000-0003-3903-0373
Sheperd S. Doeleman  https://orcid.org/0000-0002-9031-0904
Ralph P. Eatough  https://orcid.org/0000-0001-6196-4135
Heino Falcke  https://orcid.org/0000-0002-2526-6724
Vincent L. Fish  https://orcid.org/0000-0002-7128-9345
Raquel Fraga-Encinas  https://orcid.org/0000-0002-5222-1361
José L. Gómez  https://orcid.org/0000-0003-4190-7613
Peter Galison  https://orcid.org/0000-0002-6429-3872
Charles F. Gammie  https://orcid.org/0000-0001-7451-8935
Boris Georgiev  https://orcid.org/0000-0002-3586-6424
Roman Gold  https://orcid.org/0000-0003-2492-1966
Minfeng Gu (顾敏峰)  https://orcid.org/0000-0002-4455-6946

Mark Gurwell  https://orcid.org/0000-0003-0685-3621
Kazuhiro Hada  https://orcid.org/0000-0001-6906-772X
Ronald Hesper  https://orcid.org/0000-0003-1918-6098
Luis C. Ho (何子山)  https://orcid.org/0000-0001-6947-5846
Mareki Honma  https://orcid.org/0000-0003-4058-9000
Chih-Wei L. Huang  https://orcid.org/0000-0001-5641-3953
Shiro Ikeda  https://orcid.org/0000-0002-2462-1448
Sara Issaoun  https://orcid.org/0000-0002-5297-921X
David J. James  https://orcid.org/0000-0001-5160-4486
Michael Janssen  https://orcid.org/0000-0001-8685-6544
Britton Jeter  https://orcid.org/0000-0003-2847-1712
Wu Jiang (江悟)  https://orcid.org/0000-0001-7369-3539
Michael D. Johnson  https://orcid.org/0000-0002-4120-3029
Svetlana Jorstad  https://orcid.org/0000-0001-6158-1708
Taehyun Jung  https://orcid.org/0000-0001-7003-8643
Mansour Karami  https://orcid.org/0000-0001-7387-9333
Ramesh Karuppusamy  https://orcid.org/0000-0002-5307-2919
Tomohisa Kawashima  https://orcid.org/0000-0001-8527-0496
Garrett K. Keating  https://orcid.org/0000-0002-3490-146X
Mark Kettenis  https://orcid.org/0000-0002-6156-5617
Jae-Young Kim  https://orcid.org/0000-0001-8229-7183
Junhan Kim  https://orcid.org/0000-0002-4274-9373
Motoki Kino  https://orcid.org/0000-0002-2709-7338
Jun Yi Koay  https://orcid.org/0000-0002-7029-6658
Patrick M. Koch  https://orcid.org/0000-0003-2777-5861
Shoko Koyama  https://orcid.org/0000-0002-3723-3372
Michael Kramer  https://orcid.org/0000-0002-4175-2271
Carsten Kramer  https://orcid.org/0000-0002-4908-4925
Thomas P. Krichbaum  https://orcid.org/0000-0002-4892-9586
Tod R. Lauer  https://orcid.org/0000-0003-3234-7247
Sang-Sung Lee  https://orcid.org/0000-0002-6269-594X
Yan-Rong Li (李彦荣)  https://orcid.org/0000-0001-5841-9179
Zhiyuan Li (李志远)  https://orcid.org/0000-0003-0355-6437
Michael Lindqvist  https://orcid.org/0000-0002-3669-0715
Kuo Liu  https://orcid.org/0000-0002-2953-7376
Elisabetta Liuzzo  https://orcid.org/0000-0003-0995-5201
Laurent Loinard  https://orcid.org/0000-0002-5635-3345
Ru-Sen Lu (路如森)  https://orcid.org/0000-0002-7692-7967
Nicholas R. MacDonald  https://orcid.org/0000-0002-6684-8691
Jirong Mao (毛基荣)  https://orcid.org/0000-0002-7077-7195
Sera Markoff  https://orcid.org/0000-0001-9564-0876
Daniel P. Marrone  https://orcid.org/0000-0002-2367-1080
Alan P. Marscher  https://orcid.org/0000-0001-7396-3332
Iván Martí-Vidal  https://orcid.org/0000-0003-3708-9611
Lynn D. Matthews  https://orcid.org/0000-0002-3728-8082
Lia Medeiros  https://orcid.org/0000-0003-2342-6728
Karl M. Menten  https://orcid.org/0000-0001-6459-0669
Yosuke Mizuno  https://orcid.org/0000-0002-8131-6730
Izumi Mizuno  https://orcid.org/0000-0002-7210-6264
James M. Moran  https://orcid.org/0000-0002-3882-4414
Kotaro Moriyama  https://orcid.org/0000-0003-1364-3761
Monika Moscibrodzka  https://orcid.org/0000-0002-4661-6332

Cornelia Müller  https://orcid.org/0000-0002-2739-2994
Hiroshi Nagai  https://orcid.org/0000-0003-0292-3645
Neil M. Nagar  https://orcid.org/0000-0001-6920-662X
Masanori Nakamura  https://orcid.org/0000-0001-6081-2420
Ramesh Narayan  https://orcid.org/0000-0002-1919-2730
Iniyan Natarajan  https://orcid.org/0000-0001-8242-4373
Chunchong Ni  https://orcid.org/0000-0003-1361-5699
Aristeidis Noutsos  https://orcid.org/0000-0002-4151-3860
Héctor Olivares  https://orcid.org/0000-0001-6833-7580
Gisela N. Ortiz-León  https://orcid.org/0000-0002-2863-676X
Daniel C. M. Palumbo  https://orcid.org/0000-0002-7179-3816
Ue-Li Pen  https://orcid.org/0000-0003-2155-9578
Dominic W. Pesce  https://orcid.org/0000-0002-5278-9221
Oliver Porth  https://orcid.org/0000-0002-4584-2557
Ben Prather  https://orcid.org/0000-0002-0393-7734
Jorge A. Preciado-López  https://orcid.org/0000-0002-4146-0113
Hung-Yi Pu  https://orcid.org/0000-0001-9270-8812
Venkatessh Ramakrishnan  https://orcid.org/0000-0002-9248-086X
Ramprasad Rao  https://orcid.org/0000-0002-1407-7944
Alexander W. Raymond  https://orcid.org/0000-0002-5779-4767
Luciano Rezzolla  https://orcid.org/0000-0002-1330-7103
Bart Ripperda  https://orcid.org/0000-0002-7301-3908
Freek Roelofs  https://orcid.org/0000-0001-5461-3687
Eduardo Ros  https://orcid.org/0000-0001-9503-4892
Mel Rose  https://orcid.org/0000-0002-2016-8746
Alan L. Roy  https://orcid.org/0000-0002-1931-0135
Chet Ruszczyk  https://orcid.org/0000-0001-7278-9707
Benjamin R. Ryan  https://orcid.org/0000-0001-8939-4461
Kazi L. J. Rygl  https://orcid.org/0000-0003-4146-9043
David Sánchez-Arguelles  https://orcid.org/0000-0002-7344-9920
Mahito Sasada  https://orcid.org/0000-0001-5946-9960
Tuomas Savolainen  https://orcid.org/0000-0001-6214-1085
Lijing Shao  https://orcid.org/0000-0002-1334-8853
Zhiqiang Shen (沈志强)  https://orcid.org/0000-0003-3540-8746
Des Small  https://orcid.org/0000-0003-3723-5404
Bong Won Sohn  https://orcid.org/0000-0002-4148-8378
Jason SooHoo  https://orcid.org/0000-0003-1938-0720
Fumie Tazaki  https://orcid.org/0000-0003-0236-0600
Paul Tiede  https://orcid.org/0000-0003-3826-5648
Remo P. J. Tilanus  https://orcid.org/0000-0002-6514-553X
Michael Titus  https://orcid.org/0000-0002-3423-4505
Kenji Toma  https://orcid.org/0000-0002-7114-6010
Pablo Torne  https://orcid.org/0000-0001-8700-6058
Sascha Trippe  https://orcid.org/0000-0003-0465-1559
Ilse van Bemmel  https://orcid.org/0000-0001-5473-2950
Huib Jan van Langevelde  https://orcid.org/0000-0002-0230-5946
Daniel R. van Rossum  https://orcid.org/0000-0001-7772-6131
John Wardle  https://orcid.org/0000-0002-8960-2942
Jonathan Weintroub  https://orcid.org/0000-0002-4603-5204
Norbert Wex  https://orcid.org/0000-0003-4058-2837
Robert Wharton  https://orcid.org/0000-0002-7416-5209

Maciek Wielgus  https://orcid.org/0000-0002-8635-4242
George N. Wong  https://orcid.org/0000-0001-6952-2147
Qingwen Wu
(吴庆文)  https://orcid.org/0000-0003-4773-4987
André Young  https://orcid.org/0000-0003-0000-2682
Ken Young  https://orcid.org/0000-0002-3666-4920
Ziri Younsi  https://orcid.org/0000-0001-9283-1191
Feng Yuan (袁峰)  https://orcid.org/0000-0003-3564-6437
J. Anton Zensus  https://orcid.org/0000-0001-7470-3321
Guangyao Zhao  https://orcid.org/0000-0002-4417-1659
Shan-Shan Zhao  https://orcid.org/0000-0002-9774-3606
Juan-Carlos Algaba  https://orcid.org/0000-0001-6993-1696
Uwe Bach  https://orcid.org/0000-0002-7722-8412
Bradford A. Benson  https://orcid.org/0000-0002-5108-6823
Jay M. Blanchard  https://orcid.org/0000-0002-2756-395X
Iain M. Coulson  https://orcid.org/0000-0002-7316-4626
Thomas M. Crawford  https://orcid.org/0000-0001-9000-5013
Sergio A. Dzib  https://orcid.org/0000-0001-6010-6200
Wendeline B. Everett  https://orcid.org/0000-0002-5370-6651
Joseph R. Farah  https://orcid.org/0000-0003-4914-5625
Christopher H. Greer  https://orcid.org/0000-0002-9590-0508
Nils W. Halverson  https://orcid.org/0000-0003-2606-9340
Antonio Hernández-Gómez  https://orcid.org/0000-0001-7520-4305
Rubén Herrero-Illana  https://orcid.org/0000-0002-7758-8717
Atish Kamble  https://orcid.org/0000-0003-0861-5168
Ryan Keisler  https://orcid.org/0000-0002-5922-1137
Yusuke Kono  https://orcid.org/0000-0002-4187-8747
Erik M. Leitch  https://orcid.org/0000-0001-8553-9336
Kyle D. Massingill  https://orcid.org/0000-0002-0830-2033
Hugo Messias  https://orcid.org/0000-0002-2985-7994
Daniel Michalik  https://orcid.org/0000-0002-7618-6556
Andrew Nadolski  https://orcid.org/0000-0001-9479-9957
Chi H. Nguyen  https://orcid.org/0000-0001-9368-3186
Harriet Parsons  https://orcid.org/0000-0002-6327-3423
Scott N. Paine  https://orcid.org/0000-0003-4622-5857
Rurik A. Primiani  https://orcid.org/0000-0003-3910-7529
Alexandra S. Rahlin  https://orcid.org/0000-0003-3953-1776
Pim Schellart  https://orcid.org/0000-0002-8324-0880
Hotaka Shiokawa  https://orcid.org/0000-0002-8847-5275
David R. Smith  https://orcid.org/0000-0003-0692-8582
Antony A. Stark  https://orcid.org/0000-0002-2718-9996
Sjoerd T. Timmer  https://orcid.org/0000-0003-0223-9368
Nathan Whitehorn  https://orcid.org/0000-0002-3157-0007
Jan G. A. Wouterloot  https://orcid.org/0000-0002-4694-6905
Melvin Wright  https://orcid.org/0000-0002-9154-2440
Paul Yamaguchi  https://orcid.org/0000-0002-6017-8199
Lucy Ziurys  https://orcid.org/0000-0001-8205-2552

---

## The Event Horizon Telescope Collaboration,

Kazunori Akiyama[1,2,3,4] , Antxon Alberdi[5] , Walter Alef[6], Keiichi Asada[7], Rebecca Azulay[8,9,6] , Anne-Kathrin Baczko[6] ,
David Ball[10], Mislav Baloković[4,11] , John Barrett[2] , Dan Bintley[12], Lindy Blackburn[4,11] , Wilfred Boland[13],
Katherine L. Bouman[4,11,14] , Geoffrey C. Bower[15] , Michael Bremer[16], Christiaan D. Brinkerink[17] , Roger Brissenden[4,11] ,
Silke Britzen[6] , Avery E. Broderick[18,19,20] , Dominique Broguiere[16], Thomas Bronzwaer[17], Do-Young Byun[21,22] ,
John E. Carlstrom[23,24,25,26], Andrew Chael[4,11] , Chi-kwan Chan[10,27] , Shami Chatterjee[28] , Koushik Chatterjee[29],
Ming-Tang Chen[15], Yongjun Chen (陈永军)[30,31], Ilje Cho[21,22] , Pierre Christian[10,11], John E. Conway[32] , James M. Cordes[28],
Geoffrey B. Crew[2] , Yuzhu Cui[33,34] , Jordy Davelaar[17] , Mariafelicia De Laurentis[35,36,37] , Roger Deane[38,39] ,
Jessica Dempsey[12] , Gregory Desvignes[6] , Jason Dexter[40] , Sheperd S. Doeleman[4,11] , Ralph P. Eatough[6] ,
Heino Falcke[17] , Vincent L. Fish[2] , Ed Fomalont[1], Raquel Fraga-Encinas[17] , Per Friberg[12], Christian M. Fromm[36],
José L. Gómez[5] , Peter Galison[4,41,42] , Charles F. Gammie[43,44] , Roberto García[16], Olivier Gentaz[16], Boris Georgiev[19,20] ,
Ciriaco Goddi[17,45], Roman Gold[36] , Minfeng Gu (顾敏峰)[30,46] , Mark Gurwell[11] , Kazuhiro Hada[33,34] ,
Ronald Hesper[47] , Luis C. Ho (何子山)[48,49] , Paul Ho[7], Mareki Honma[33,34] , Chih-Wei L. Huang[7] , Lei Huang (黄磊)[30,46] ,
David H. Hughes[50], Shiro Ikeda[3,51,52,53] , Makoto Inoue[7], Sara Issaoun[17] , David J. James[4,11] , Buell T. Jannuzi[10],
Michael Janssen[17] , Britton Jeter[19,20] , Wu Jiang (江悟)[30] , Michael D. Johnson[4,11] , Svetlana Jorstad[54,55] ,
Taehyun Jung[21,22] , Mansour Karami[18,19] , Ramesh Karuppusamy[6] , Tomohisa Kawashima[3] , Garrett K. Keating[11] ,
Mark Kettenis[56] , Jae-Young Kim[6] , Junhan Kim[10] , Jongsoo Kim[21], Motoki Kino[3,57] , Jun Yi Koay[7] ,
Patrick M. Koch[7] , Shoko Koyama[7] , Michael Kramer[6] , Carsten Kramer[16] , Thomas P. Krichbaum[6] , Cheng-Yu Kuo[58],
Tod R. Lauer[59] , Sang-Sung Lee[21] , Yan-Rong Li (李彦荣)[60] , Zhiyuan Li (李志远)[61,62] , Michael Lindqvist[32] ,
Kuo Liu[6] , Elisabetta Liuzzo[63] , Wen-Ping Lo[7,64] , Andrei P. Lobanov[6], Laurent Loinard[65,66] , Colin Lonsdale[2], Ru-Sen Lu
(路如森)[6,30] , Nicholas R. MacDonald[6] , Jirong Mao (毛基荣)[67,68,69] , Sera Markoff[29,70] , Daniel P. Marrone[10] ,

Alan P. Marscher[54] , Iván Martí-Vidal[32,71] , Satoki Matsushita[7], Lynn D. Matthews[2] , Lia Medeiros[10,72] ,
Karl M. Menten[6] , Yosuke Mizuno[36] , Izumi Mizuno[12] , James M. Moran[4,11] , Kotaro Moriyama[2,33] ,
Monika Moscibrodzka[17] , Cornelia Müller[6,17] , Hiroshi Nagai[3,34] , Neil M. Nagar[73] , Masanori Nakamura[7] ,
Ramesh Narayan[4,11] , Gopal Narayanan[74] , Iniyan Natarajan[39] , Roberto Neri[16], Chunchong Ni[19,20] , Aristeidis Noutsos[6] ,
Hiroki Okino[33,75], Héctor Olivares[36] , Gisela N. Ortiz-León[33] , Tomoaki Oyama[33], Feryal Özel[10], Daniel C. M. Palumbo[4,11] ,
Nimesh Patel[11], Ue-Li Pen[18,76,77,78] , Dominic W. Pesce[4,11] , Vincent Piétu[16], Richard Plambeck[79], Aleksandar PopStefanija[74],
Oliver Porth[29,36] , Ben Prather[43] , Jorge A. Preciado-López[18] , Dimitrios Psaltis[10], Hung-Yi Pu[18] ,
Venkatessh Ramakrishnan[73] , Ramprasad Rao[15] , Mark G. Rawlings[12], Alexander W. Raymond[4,11] , Luciano Rezzolla[36] ,
Bart Ripperda[36] , Freek Roelofs[17] , Alan Rogers[2], Eduardo Ros[6] , Mel Rose[10] , Arash Roshanineshat[10], Helge Rottmann[6],
Alan L. Roy[6] , Chet Ruszczyk[2] , Benjamin R. Ryan[80,81] , Kazi L. J. Rygl[63] , Salvador Sánchez[82],
David Sánchez-Arguelles[50,83] , Mahito Sasada[33,84] , Tuomas Savolainen[6,85,86] , F. Peter Schloerb[74], Karl-Friedrich Schuster[16],
Lijing Shao[6,49] , Zhiqiang Shen (沈志强)[30,31] , Des Small[56] , Bong Won Sohn[21,22,87] , Jason SooHoo[2] , Fumie Tazaki[33] ,
Paul Tiede[19,20] , Remo P. J. Tilanus[17,45,88] , Michael Titus[2] , Kenji Toma[89,90] , Pablo Torne[6,82] , Tyler Trent[10],
Sascha Trippe[91] , Shuichiro Tsuda[33], Ilse van Bemmel[56] , Huib Jan van Langevelde[56,92] , Daniel R. van Rossum[17] ,
Jan Wagner[6], John Wardle[93] , Jonathan Weintroub[4,11] , Norbert Wex[6], Robert Wharton[6] , Maciek Wielgus[4,11] ,
George N. Wong[43] , Qingwen Wu (吴庆文)[94] , André Young[17] , Ken Young[11] , Ziri Younsi[95,36] , Feng Yuan
(袁峰)[30,46,96] , Ye-Fei Yuan (袁业飞)[97] , J. Anton Zensus[6] , Guangyao Zhao[73] , Shan-Shan Zhao[17,61] , Ziyan Zhu[42],
Juan-Carlos Algaba[7,98] , Alexander Allardi[99], Rodrigo Amestica[100], Uwe Bach[6] , Christopher Beaudoin[2],
Bradford A. Benson[23,24,26] , Ryan Berthold[12], Jay M. Blanchard[56,73] , Ray Blundell[11], Sandra Bustamente[101], Roger Cappallo[2],
Edgar Castillo-Domínguez[101,102], Chih-Cheng Chang[7,103], Shu-Hao Chang[7], Song-Chu Chang[103], Chung-Chen Chen[7],
Ryan Chilson[15], Tim C. Chuter[12], Rodrigo Córdova Rosado[4,11] , Iain M. Coulson[12] , Thomas M. Crawford[24] ,
Joseph Crowley[104], John David[82], Mark Derome[2], Matthew Dexter[105], Sven Dornbusch[6], Kevin A. Dudevoir[135],
Sergio A. Dzib[6] , Chris Eckert[2], Neal R. Erickson[74], Wendeline B. Everett[106] , Aaron Faber[107], Joseph R. Farah[4,11,108] ,
Vernon Fath[74], Thomas W. Folkers[74], David C. Forbes[10], Robert Freund[101,102], Arturo I. Gómez-Ruiz[101,102], David Gale[101],
Feng Gao[30,40], Gertie Geertsema[109], David A. Graham[6], Christopher H. Greer[10], Ronald Grosslein[74], Frédéric Gueth[16],
Nils W. Halverson[110] , Chih-Chiang Han[7], Kuo-Chang Han[103], Jinchi Hao[103], Yutaka Hasegawa[111], Jason W. Henning[112,23] ,
Antonio Hernández-Gómez[65,113] , Rubén Herrero-Illana[114] , Stefan Heyminck[6], Akihiko Hirota[3,115], James Hoge[12],
Yau-De Huang[7], C. M. Violette Impellizzeri[115,116], Homin Jiang[7], Atish Kamble[4,11] , Ryan Keisler[25] , Kimihiro Kimura[111],
Yusuke Kono[12] , Derek Kubo[117], John Kuroda[12], Richard Lacasse[100], Robert A. Laing[118] , Erik M. Leitch[23] , Chao-Te Li[7],
Lupin C.-C. Lin[7,119] , Ching-Tang Liu[103], Kuan-Yu Liu[7], Li-Ming Lu[103], Ralph G. Marson[120], Pierre L. Martin-Cocher[7],
Kyle D. Massingill[10], Callie Matulonis[12], Martin P. McColl[10], Stephen R. McWhirter[2], Hugo Messias[114,115] ,
Zheng Meyer-Zhao[7,121], Daniel Michalik[122,24] , Alfredo Montaña[101,102], William Montgomerie[12], Matias Mora-Klein[100],
Dirk Muders[6], Andrew Nadolski[44] , Santiago Navarro[82], Chi H. Nguyen[10,123] , Hiroaki Nishioka[7], Timothy Norton[11],
George Nystrom[15], Hideo Ogawa[111], Peter Oshiro[15], Tomoaki Oyama[124], Stephen Padin[23,24], Harriet Parsons[12] ,
Scott N. Paine[11] , Juan Peñalver[82], Neil M. Phillips[114,115], Michael Poirier[2], Nicolas Pradel[7], Rurik A. Primiani[125] ,
Philippe A. Raffin[15], Alexandra S. Rahlin[23,126] , George Reiland[10], Christopher Risacher[16], Ignacio Ruiz[82],
Alejandro F. Sáez-Madaín[100,115] , Remi Sassella[16], Pim Schellart[17,127] , Paul Shaw[7], Kevin M. Silva[11], Hotaka Shiokawa[11] ,
David R. Smith[128,129], William Snow[15], Kamal Souccar[74], Don Sousa[2], T. K. Sridharan[11], Ranjani Srinivasan[15],
William Stahm[12], Antony A. Stark[11] , Kyle Story[130], Sjoerd T. Timmer[17] , Laura Vertatschitsch[11,125], Craig Walther[12],
Ta-Shun Wei[7], Nathan Whitehorn[131] , Alan R. Whitney[2], David P. Woody[132], Jan G. A. Wouterloot[12] , Melvin Wright[133],
Paul Yamaguchi[11], Chen-Yu Yu[7], Milagros Zeballos[101,134], and Lucy Ziurys[10]

[1] National Radio Astronomy Observatory, 520 Edgemont Rd, Charlottesville, VA 22903, USA
[2] Massachusetts Institute of Technology Haystack Observatory, 99 Millstone Road, Westford, MA 01886, USA
[3] National Astronomical Observatory of Japan, 2-21-1 Osawa, Mitaka, Tokyo 181-8588, Japan
[4] Black Hole Initiative at Harvard University, 20 Garden Street, Cambridge, MA 02138, USA
[5] Instituto de Astrofísica de Andalucía-CSIC, Glorieta de la Astronomía s/n, E-18008 Granada, Spain
[6] Max-Planck-Institut für Radioastronomie, Auf dem Hügel 69, D-53121 Bonn, Germany
[7] Institute of Astronomy and Astrophysics, Academia Sinica, 11F of Astronomy-Mathematics Building,
  AS/NTU No. 1, Sec. 4, Roosevelt Rd, Taipei 10617, Taiwan, R.O.C.
[8] Departament d'Astronomia i Astrofísica, Universitat de València, C. Dr. Moliner 50, E-46100 Burjassot, València, Spain
[9] Observatori Astronòmic, Universitat de València, C. Catedrático José Beltrán 2, E-46980 Paterna, València, Spain
[10] Steward Observatory and Department of Astronomy, University of Arizona, 933 N. Cherry Ave., Tucson, AZ 85721, USA
[11] Center for Astrophysics | Harvard & Smithsonian, 60 Garden Street, Cambridge, MA 02138, USA
[12] East Asian Observatory, 660 N. A'ohoku Place, Hilo, HI 96720, USA
[13] Nederlandse Onderzoekschool voor Astronomie (NOVA), PO Box 9513, 2300 RA Leiden, The Netherlands
[14] California Institute of Technology, 1200 East California Boulevard, Pasadena, CA 91125, USA
[15] Institute of Astronomy and Astrophysics, Academia Sinica, 645 N. A'ohoku Place, Hilo, HI 96720, USA
[16] Institut de Radioastronomie Millimétrique, 300 rue de la Piscine, F-38406 Saint Martin d'Hères, France
[17] Department of Astrophysics, Institute for Mathematics, Astrophysics and Particle Physics (IMAPP), Radboud University,
  P.O. Box 9010, 6500 GL Nijmegen, The Netherlands

[18] Perimeter Institute for Theoretical Physics, 31 Caroline Street North, Waterloo, ON, N2L 2Y5, Canada

[19] Department of Physics and Astronomy, University of Waterloo, 200 University Avenue West, Waterloo, ON, N2L 3G1, Canada

[20] Waterloo Centre for Astrophysics, University of Waterloo, Waterloo, ON N2L 3G1 Canada

[21] Korea Astronomy and Space Science Institute, Daedeok-daero 776, Yuseong-gu, Daejeon 34055, Republic of Korea

[22] University of Science and Technology, Gajeong-ro 217, Yuseong-gu, Daejeon 34113, Republic of Korea

[23] Kavli Institute for Cosmological Physics, University of Chicago, 5640 South Ellis Avenue, Chicago, IL 60637, USA

[24] Department of Astronomy and Astrophysics, University of Chicago, 5640 South Ellis Avenue, Chicago, IL 60637, USA

[25] Department of Physics, University of Chicago, 5640 South Ellis Avenue, Chicago, IL 60637, USA

[26] Enrico Fermi Institute, University of Chicago, 5640 South Ellis Avenue, Chicago, IL 60637, USA

[27] Data Science Institute, University of Arizona, 1230 N. Cherry Ave., Tucson, AZ 85721, USA

[28] Cornell Center for Astrophysics and Planetary Science, Cornell University, Ithaca, NY 14853, USA

[29] Anton Pannekoek Institute for Astronomy, University of Amsterdam, Science Park 904, 1098 XH, Amsterdam, The Netherlands

[30] Shanghai Astronomical Observatory, Chinese Academy of Sciences, 80 Nandan Road, Shanghai 200030, People's Republic of China

[31] Key Laboratory of Radio Astronomy, Chinese Academy of Sciences, Nanjing 210008, People's Republic of China

[32] Department of Space, Earth and Environment, Chalmers University of Technology, Onsala Space Observatory, SE-43992 Onsala, Sweden

[33] Mizusawa VLBI Observatory, National Astronomical Observatory of Japan, 2-12 Hoshigaoka, Mizusawa, Oshu, Iwate 023-0861, Japan

[34] Department of Astronomical Science, The Graduate University for Advanced Studies (SOKENDAI), 2-21-1 Osawa, Mitaka, Tokyo 181-8588, Japan

[35] Dipartimento di Fisica "E. Pancini", Universitá di Napoli "Federico II", Compl. Univ. di Monte S. Angelo, Edificio G, Via Cinthia, I-80126, Napoli, Italy

[36] Institut für Theoretische Physik, Goethe-Universität Frankfurt, Max-von-Laue-Straße 1, D-60438 Frankfurt am Main, Germany

[37] INFN Sez. di Napoli, Compl. Univ. di Monte S. Angelo, Edificio G, Via Cinthia, I-80126, Napoli, Italy

[38] Department of Physics, University of Pretoria, Lynnwood Road, Hatfield, Pretoria 0083, South Africa

[39] Centre for Radio Astronomy Techniques and Technologies, Department of Physics and Electronics, Rhodes University, Grahamstown 6140, South Africa

[40] Max-Planck-Institut für Radioastronomie, Auf dem Hügel 69, D-53121 Bonn, Germany

[40] Max-Planck-Institut für extraterrestrische Physik, Giessenbachstr. 1, D-85748 Garching, Germany

[41] Department of History of Science, Harvard University, Cambridge, MA 02138, USA

[42] Department of Physics, Harvard University, Cambridge, MA 02138, USA

[43] Department of Physics, University of Illinois, 1110 West Green St, Urbana, IL 61801, USA

[44] Department of Astronomy, University of Illinois at Urbana-Champaign, 1002 West Green Street, Urbana, IL 61801, USA

[45] Leiden Observatory—Allegro, Leiden University, P.O. Box 9513, 2300 RA Leiden, The Netherlands

[46] Key Laboratory for Research in Galaxies and Cosmology, Chinese Academy of Sciences, Shanghai 200030, People's Republic of China

[47] NOVA Sub-mm Instrumentation Group, Kapteyn Astronomical Institute, University of Groningen, Landleven 12, 9747 AD Groningen, The Netherlands

[48] Department of Astronomy, School of Physics, Peking University, Beijing 100871, People's Republic of China

[49] Kavli Institute for Astronomy and Astrophysics, Peking University, Beijing 100871, People's Republic of China

[50] Instituto Nacional de Astrofísica, Óptica y Electrónica. Apartado Postal 51 y 216, 72000. Puebla Pue., México

[51] The Institute of Statistical Mathematics, 10-3 Midori-cho, Tachikawa, Tokyo, 190-8562, Japan

[52] Department of Statistical Science, The Graduate University for Advanced Studies (SOKENDAI), 10-3 Midori-cho, Tachikawa, Tokyo 190-8562, Japan

[53] Kavli Institute for the Physics and Mathematics of the Universe, The University of Tokyo, 5-1-5 Kashiwanoha, Kashiwa, 277-8583, Japan

[54] Institute for Astrophysical Research, Boston University, 725 Commonwealth Ave., Boston, MA 02215, USA

[55] Astronomical Institute, St. Petersburg University, Universitetskij pr., 28, Petrodvorets,198504 St.Petersburg, Russia

[56] Joint Institute for VLBI ERIC (JIVE), Oude Hoogeveensedijk 4, 7991 PD Dwingeloo, The Netherlands

[57] Kogakuin University of Technology & Engineering, Academic Support Center, 2665-1 Nakano, Hachioji, Tokyo 192-0015, Japan

[58] Physics Department, National Sun Yat-Sen University, No. 70, Lien-Hai Rd, Kaosiung City 80424, Taiwan, R.O.C

[59] National Optical Astronomy Observatory, 950 North Cherry Ave., Tucson, AZ 85719, USA

[60] Key Laboratory for Particle Astrophysics, Institute of High Energy Physics, Chinese Academy of Sciences, 19B Yuquan Road, Shijingshan District, Beijing, People's Republic of China

[61] School of Astronomy and Space Science, Nanjing University, Nanjing 210023, People's Republic of China

[62] Key Laboratory of Modern Astronomy and Astrophysics, Nanjing University, Nanjing 210023, People's Republic of China

[63] Italian ALMA Regional Centre, INAF-Istituto di Radioastronomia, Via P. Gobetti 101, I-40129 Bologna, Italy

[64] Department of Physics, National Taiwan University, No.1, Sect.4, Roosevelt Rd., Taipei 10617, Taiwan, R.O.C

[65] Instituto de Radioastronomía y Astrofísica, Universidad Nacional Autónoma de México, Morelia 58089, México

[66] Instituto de Astronomía, Universidad Nacional Autónoma de México, CdMx 04510, México

[67] Yunnan Observatories, Chinese Academy of Sciences, 650011 Kunming, Yunnan Province, People's Republic of China

[68] Center for Astronomical Mega-Science, Chinese Academy of Sciences, 20A Datun Road, Chaoyang District, Beijing, 100012, People's Republic of China

[69] Key Laboratory for the Structure and Evolution of Celestial Objects, Chinese Academy of Sciences, 650011 Kunming, People's Republic of China

[70] Gravitation Astroparticle Physics Amsterdam (GRAPPA) Institute, University of Amsterdam, Science Park 904, 1098 XH Amsterdam, The Netherlands

[71] Centro Astronómico de Yebes (IGN), Apartado 148, E-19180 Yebes, Spain

[72] Department of Physics, Broida Hall, University of California Santa Barbara, Santa Barbara, CA 93106, USA

[73] Astronomy Department, Universidad de Concepción, Casilla 160-C, Concepción, Chile

[74] Department of Astronomy, University of Massachusetts, 01003, Amherst, MA, USA

[75] Department of Astronomy, Graduate School of Science, The University of Tokyo, 7-3-1 Hongo, Bunkyo-ku, Tokyo 113-0033, Japan

[76] Canadian Institute for Theoretical Astrophysics, University of Toronto, 60 St. George Street, Toronto, ON M5S 3H8, Canada

[77] Dunlap Institute for Astronomy and Astrophysics, University of Toronto, 50 St. George Street, Toronto, ON M5S 3H4, Canada

[78] Canadian Institute for Advanced Research, 180 Dundas St West, Toronto, ON M5G 1Z8, Canada

[79] Radio Astronomy Laboratory, University of California, Berkeley, CA 94720, USA

[80] CCS-2, Los Alamos National Laboratory, P.O. Box 1663, Los Alamos, NM 87545, USA

[81] Center for Theoretical Astrophysics, Los Alamos National Laboratory, Los Alamos, NM, 87545, USA

[82] Instituto de Radioastronomía Milimétrica, IRAM, Avenida Divina Pastora 7, Local 20, E-18012, Granada, Spain

[83] Consejo Nacional de Ciencia y Tecnología, Av. Insurgentes Sur 1582, 03940, Ciudad de México, México

[84] Hiroshima Astrophysical Science Center, Hiroshima University, 1-3-1 Kagamiyama, Higashi-Hiroshima, Hiroshima 739-8526, Japan

[85] Aalto University Department of Electronics and Nanoengineering, PL 15500, FI-00076 Aalto, Finland

[86] Aalto University Metsähovi Radio Observatory, Metsähovintie 114, FI-02540 Kylmälä, Finland

[87] Department of Astronomy, Yonsei University, Yonsei-ro 50, Seodaemun-gu, 03722 Seoul, Republic of Korea

[88] Netherlands Organisation for Scientific Research (NWO), Postbus 93138, 2509 AC Den Haag, The Netherlands

[89] Frontier Research Institute for Interdisciplinary Sciences, Tohoku University, Sendai 980-8578, Japan

[90] Astronomical Institute, Tohoku University, Sendai 980-8578, Japan

[91] Department of Physics and Astronomy, Seoul National University, Gwanak-gu, Seoul 08826, Republic of Korea

[92] Leiden Observatory, Leiden University, Postbus 2300, 9513 RA Leiden, The Netherlands

[93] Physics Department, Brandeis University, 415 South Street, Waltham, MA 02453, USA

[94] School of Physics, Huazhong University of Science and Technology, Wuhan, Hubei, 430074, People's Republic of China

[95] Mullard Space Science Laboratory, University College London, Holmbury St. Mary, Dorking, Surrey, RH5 6NT, UK

[96] School of Astronomy and Space Sciences, University of Chinese Academy of Sciences, No. 19A Yuquan Road, Beijing 100049, People's Republic of China

[97] Astronomy Department, University of Science and Technology of China, Hefei 230026, People's Republic of China

[98] Department of Physics, Faculty of Science, University of Malaya, 50603 Kuala Lumpur, Malaysia

[99] University of Vermont, Burlington, VT 05405, USA

[100] National Radio Astronomy Observatory, NRAO Technology Center, 1180 Boxwood Estate Road, Charlottesville, VA 22903, USA

[101] Instituto Nacional de Astrofísica, Óptica y Electrónica, Luis Enrique Erro 1, Tonantzintla, Puebla, C.P. 72840, Mexico

[102] Consejo Nacional de Ciencia y Tecnología, Av. Insurgentes Sur 1582, Col. Crédito Constructor, CDMX, C.P. 03940, Mexico

[103] National Chung-Shan Institute of Science and Technology, No.566, Ln. 134, Longyuan Rd., Longtan Dist., Taoyuan City 325, Taiwan, R.O.C.

[104] MIT Haystack Observatory, 99 Millstone Road, Westford, MA 01886, USA

[105] Dept. of Astronomy, Univ. of California Berkeley, 501 Campbell, Berkeley, CA 94720, USA

[106] CASA, Department of Astrophysical and Planetary Sciences, University of Colorado, Boulder, CO 80309, USA

[107] Western University, 1151 Richmond Street, London, Ontario, N6A 3K7, Canada

[108] University of Massachusetts Boston, 100 William T, Morrissey Blvd, Boston, MA 02125, USA

[109] Research and Development Weather and Climate Models, Royal Netherlands Meteorological Institute, Utrechtseweg 297, 3731 GA, De Bilt, The Netherlands

[110] Department of Astrophysical and Planetary Sciences and Department of Physics, University of Colorado, Boulder, CO 80309, USA

[111] Osaka Prefecture University, Gakuencyou Sakai Osaka, Sakai 599-8531, Kinki, Japan

[112] High Energy Physics Division, Argonne National Laboratory, 9700 S. Cass Avenue, Argonne, IL 60439, USA

[113] IRAP, Université de Toulouse, CNRS, UPS, CNES, Toulouse, France

[114] European Southern Observatory, Alonso de Córdova 3107, Vitacura, Casilla 19001, Santiago de Chile, Chile

[115] Joint ALMA Observatory, Alonso de Córdova 3107, Vitacura 763-0355, Santiago de Chile, Chile

[116] National Radio Astronomy Observatory, 520 Edgemont Road, Charlottesville, VA 22903, USA

[117] ASIAA Hilo Office, 645 N. A'ohoku Place, University Park, Hilo, HI 96720, USA

[118] Square Kilometre Array Organisation, Jodrell Bank Observatory, Lower Withington, Macclesfield, Cheshire SK11 9DL, UK

[119] Department of Physics, Ulsan National Institute of Science and Technology, Ulsan, 44919, Republic of Korea

[120] National Radio Astronomy Observatory, P.O. Box O, Socorro, NM 87801, USA

[121] ASTRON, Oude Hoogeveensedijk 4, 7991 PD Dwingeloo, The Netherlands

[122] Science Support Office, Directorate of Science, European Space Research and Technology Centre (ESA/ESTEC), Keplerlaan 1, 2201 AZ Noordwijk, The Netherlands

[123] Center for Detectors, School of Physics and Astronomy, Rochester Institute of Technology, 1 Lomb Memorial Drive, Rochester, NY 14623, USA

[125] Mizusawa VLBI Observatory, National Astronomical Observatory of Japan, Ohshu, Iwate 023-0861, Japan

[125] Systems & Technology Research, 600 West Cummings Park, Woburn, MA 01801, USA

[126] Fermi National Accelerator Laboratory, PO Box 500, Batavia IL 60510, USA

[127] Department of Astrophysical Sciences, Princeton University, Princeton, NJ 08544, USA

[128] MERLAB, 357 S. Candler St., Decatur, GA 30030, USA

[129] GWW School of Mechanical Engineering, Georgia Institute of Technology, 771 Ferst Dr., Atlanta, GA 30332, USA

[130] Kavli Institute for Particle Astrophysics and Cosmology, Stanford University, 452 Lomita Mall, Stanford, CA 94305, USA

[131] Dept. of Physics and Astronomy, UCLA, Los Angeles, CA 90095, USA

[132] Owens Valley Radio Observatory, California Institute of Technology, Big Pine, CA 93513, USA

[133] Dept. of Astronomy, Radio Astronomy Laboratory, Univ. of California Berkeley, 601 Campbell, Berkeley, CA 94720, USA

[134] Universidad de las Américas Puebla, Sta. Catarina Mártir S/N, San Andrés Cholula, Puebla, C.P. 72810, Mexico

[135] Deceased.